\documentclass[reprint, amsmath,amssymb,aps]{revtex4-1}
\def\q{{}^o\!q}
\def\eb{\!\bar{e}}
\def\wb{\!\bar{\omega}}
\def\c{\bar{c}}
\def\p{\bar{p}}
\def\k{\bar{k}}
\def\pa{\partial}
\def\nn{\nonumber}
\def\be{\begin{equation}}
\def\ee{\end{equation}}
\def\ba{\begin{eqnarray}}
\def\ea{\end{eqnarray}}
\begin{document}
\title{Cosmological perturbations with inverse-volume corrections \\in loop quantum cosmology}

\author{Yu Han}
 \email{hanyu@xynu.edu.cn}
 \affiliation{College of Physics and Electrical Engineering, Xinyang Normal University, 464000 Xinyang, China}
\date{\today}

\begin{abstract}
Although the cosmological perturbations with inverse-volume corrections from loop quantum cosmology have been studied using the anomaly-free algebra approach in much of the literature,  there still remains an important issue that some counterterms in the perturbed constraints cannot be uniquely fixed on the spatially flat Friedmann-Robterson-Walker background, which causes ambiguities in the perturbation equations. In this paper, we show that this problem can be overcome by extending the anomaly-free algebra to the spatially closed Friedmann-Robterson-Walker background. We find that a consistent deformed algebra can be obtained in the spatially closed case,  and each counter term can be uniquely fixed in terms of the inverse-volume correction functions; then, by taking the large $r_o$ limit, we recover the anomaly-free Hamiltonian on the spatially flat background. Using this Hamiltonian we obtain the gauge invariant cosmological perturbations for scalar, vector and tensor modes in the spatially flat case. Moreover, we also derive the quantum-corrected Mukhanov equations, from which the scalar and tensor spectral indices with inverse-volume corrections are given. Some key cosmological perturbation equations obtained in this paper are different from those in previous literature.
\end{abstract}
\maketitle
\section{Introduction}
Among different attempts to search for the theory of quantum gravity, loop quantum gravity (LQG) is a representative of those theories which believe the fundamental nature of gravity is spacetime geometry and the core issue of quantum gravity is to find a suitable way to quantize the geometry. One of the important predictions of LQG is that the spatial geometry is discrete at the Planck scale. The early Universe provides a natural testing field for such predictions. However, due to the complexity of the full theory, it is extremely difficult to directly study the cosmology from LQG. An expedient program called loop quantum cosmology (LQC) which applies the quantization techniques of LQG to the symmetry reduced model of cosmology is currently widely used.

Despite a symmetry reduced quantized model, LQC captures many important features of LQG. Therefore, with some justification, many predictions of LQC are believed to reflect the genuine quantum gravity effect from the full theory in a unrefined manner. On the effective level, two characteristic corrections, the holonomy corrections and the inverse-volume corrections, which are initially proposed in LQG, have been extensively investigated in LQC \cite{Ashtekar:2011,Bojowald:2006}. Roughly speaking, the regimes in which the two corrections dominate are different because the inverse-volume corrections depend on the quantum gravity scale whereas the holonomy corrections mainly depend on the higher powers of extrinsic curvature; hence, the two corrections can often be separately studied \cite{Barrau:2014}. For observational interest, under certain parametrization the inverse-volume correction can give rise to much stronger quantum gravity effects than the holonomy correction during the slow-roll inflation even though the energy density is much smaller than the Planck energy scale \cite{Bojowald:2011b}. To probe more detectable quantum gravity effects, the cosmological perturbations in the context of LQC have been intensely studied during the past decade. Currently, three different approaches to treat the effective perturbations have been proposed and are separately called the ``anomaly-free constraint algebra" \cite{Bojowald:2008a,Cailleteau:2012}, ``hybrid models" \cite{Martin-Benito:2008,Garay:2010} and ``dressed metric" \cite{Agullo:2012,Agullo:2013}. Each approach has its own advantage and can also be questioned in some aspects. In this paper, we follow the anomaly-free constraint algebra approach.

 In isotropic and homogenous LQC, the diffeomorphism constraint vanishes, and the Hamiltonian constraint trivially commutes with itself. However, the algebra becomes much more involved after including the perturbations. If the perturbations are included in the constraints in the classical form, calculations show that the Poisson brackets are no longer closed and the anomalies appear, which means the covariance of the theory is broken on the effective level.  A natural way to cancel these anomalies is to add in the perturbed constraint some undetermined counterterms; then, by requiring a closed constraint algebra and asking the counterterms to vanish in the classical limit we can fix their expressions. Once the anomaly-free algebra is obtained, it means we have found a consistent effective theory.

 The anomaly-free algebra approach was first developed in Ref. \cite{Bojowald:2008a} for inverse-volume corrections. The cosmological gauge invariant perturbations and the power spectra with quantum corrections were derived  in Refs. \cite{Bojowald:2008b,Bojowald:2010}, and the observational constraints with minimally coupled scalar matter were studied in Refs. \cite{Bojowald:2011a,Bojowald:2011b}. For holonomy corrections, this approach was first applied to the scalar perturbations in Ref. \cite{Cailleteau:2011a} and to all kinds of perturbations in Refs. \cite{Cailleteau:2012a,Han:2017}. The anomaly-free algebra including both inverse-volume and holonomy corrections is studied in Ref. \cite{Cailleteau:2013}. Besides the cosmological models, this approach has also been tentatively applied to many other models with different symmetries, such as the spherical symmetry \cite{Bojowald:2015a}, Gowdy cosmology \cite{Bojowald:2015b} and two-dimensional dilaton gravity \cite{Bojowald:2016}.
 One of the most important results of the anomaly-free approach is that the classical constraint algebra is deformed by quantum corrections;
 i.e. the Poisson bracket between two Hamiltonian constraints becomes
 \begin{equation}
 \{H[N_1],H[N_2]\}=D_a[\Omega(N_1\nabla^aN_2-N_2\nabla^aN_1)] \label{DefHHbrac}
 \end{equation}
 where $\Omega$ is a phase-space function, which is different from the classical case where $\Omega=1$, and all the other brackets remain unchanged; thus, the covariance of the theory is retained. The explicit form of function $\Omega$ is closely related to the specific quantum correction implemented.  On the spatially flat FRW background, previous results tell us that the explicit form of $\Omega$ cannot be uniquely determined not only due to the quantization ambiguities in the corrections but also because the requirement of anomaly freedom is not strong enough to fix all the counterterms, some of which remain free functions and appear in the expression of $\Omega$.
 To avoid the unwanted free functions and simplify the calculations, in Ref. \cite{Bojowald:2008a}, some counterterms were presumed to vanish (which was also presumed in Ref. \cite{Cailleteau:2011a} for holonomy corrections) and only linear terms in the anomalies were considered when calculating the Poisson algebra.  In Ref. \cite{Cailleteau:2013}, the authors reconsidered the issue by including all the possible counterterms and solved the nonlinear anomalies exactly, they found that the counterterms can be fixed up to a free function $f_1[p]$, which also shows up in $\Omega$, and thus the ambiguities in the theory still cannot be avoided.

 In this paper, we try to solve this problem from a different perspective. To be concrete, in Secs.II and III we extend the anomaly-free approach for inverse-volume corrections to spatially closed Friedmann-Robterson-Walker (FRW) background and derive the expression of each counterterm. Since in the large $r_o$ limit the constraints on the spatially closed background can reduce to the one on the spatially flat background, the anomaly-free Hamiltonian with determined counterterms on the spatially flat background can then be directly obtained. With the help of this Hamiltonian, in Sec. IV, we derive the gauge invariant cosmological perturbations of scalar, vector and tensor modes in spatially flat case. The Mukhanov equations for density and tensor perturbations will also be given there. In Sec. V, we derive the quantum-corrected scalar and tensor spectral indices. In the last section, we make some remarks.

\section{Constraints on spatially closed FRW background}
This section is divided into two parts. We first review the basic variables and background constraint of the spatially closed FRW model, and in the second part, we derive the linearly perturbed constraints in terms of connection variables.
\subsection{Background Hamiltonian constraint}
In this part, we summarize the classical theory of the spatially closed FRW model developed in Refs. \cite{Ashtekar:2006,Szulc:2006} using connection variables. To facilitate a more clear comparison with the spatially flat case, we will use notations slightly different from those in Refs. \cite{Ashtekar:2006,Szulc:2006}. In the spatially closed case,
the space manifold $\Sigma$ has the topology of a 3-sphere, $\mathbb{S}^3$, on which we have an isotropic and homogenous, left invariant coframe\, $\wb^i_a$, and its dual frame\, $\eb^a_i$, the fiducial metric is given by
\be
\q_{ab}\equiv \wb_a^i\, \wb_b^j\, k_{ij},
\ee
where $k_{ij}$ denotes the Cartan-Killing metric on su(2). We denote the radius of the 3-sphere $\Sigma$ as $r_o$ with respect to the fiducial metric. In order to better
clarify the relation with the spatially flat case, the radius $r_o$ here can correspond to an arbitrary real number, which is different from Ref. \cite{Ashtekar:2006} wherein the authors fixed $r_o=2$. The isotropic, homogenous connections and densitized triads can be separately parametrized by $\bar{c}$ and $\bar{p}$,
\ba
\bar{A}_a^i=\c\,\,\wb_a^i,\quad\quad \bar{E}^a_i=\bar{p}\,\sqrt{\q}\,\,\eb^a_i,
\ea
satisfying
\be
\{\c,\p\}=\frac{\kappa\gamma}{3V_0},\qquad \kappa=8\pi G, \label{Poissoncp}
\ee
 where $\q\equiv\det(\q_{ab})$ and $\gamma$ is the Barbero-Immirzi parameter and $V_0$ is the volume of some cell $\mathcal{C}$ characteristic of the regime considered (which is not necessarily the 3-sphere $\Sigma$) with respect to the fiducial metric. Straightforward computation shows that the spin connection is given by
\be
\bar{\Gamma}^i_a=\frac{1}{r_o}\,\wb^i_a. \label{cgamma}
\ee
The extrinsic curvature $\bar{K}_a^i\equiv\frac{1}{\gamma}(\bar{A}_a^i-\bar{\Gamma}^i_a)$ becomes
\be
 \bar{K}_a^i=\frac{1}{\gamma}(\c-\frac{1}{r_o})\,\,\wb^i_a=:\k\,\,\wb^i_a \label{ckai}
\ee
with $\{\k,\p\}=\frac{\kappa}{3V_0}$; i.e.
 the variables $\k$ and $\p$ now depend on the fiducial volume $V_0$. Obviously the spin connection vanishes as $r_o$ becomes large enough and the Poisson bracket (\ref{Poissoncp}) will be identical with the one in the spatially flat case if we set $V_0$ equal to the fiducial volume ${\cal V}_0$ used there.

 In the homogeneous and isotropic cases, the background Gauss and diffeomorphism constraints vanish; hence, we only need to consider the background Hamiltonian constraint. The  gravitational part of the Hamiltonian constraint is
\be
\textbf{H}_{g}[N]=\frac{1}{2\kappa}\int_\mathcal{C} d^3x N\mathcal{H}_{g},
\ee
with
\be
\mathcal{H}_{g}=\epsilon^{ij}{}_{k}\frac{E^a_iE^b_j}{\sqrt{|\det E|}}[2\pa_a\Gamma_b^k+\epsilon_{mn}{}^{k}(\Gamma^m_a\Gamma^n_b-K^m_aK^n_b)].\label{mathH}
\ee

Substituting (\ref{cgamma}) and (\ref{ckai}) into (\ref{mathH}), we easily find the background gravitational Hamiltonian constraint is expressed by
\ba
\bar{\textbf{H}}_{g}[\bar{N}]&=&\frac{1}{2\kappa}\int_\mathcal{C} d^3x\bar{N}\mathcal{H}^{(0)}_g\nn\\
&=&\frac{1}{2\kappa}\int_\mathcal{C} d^3x\bar{N}\sqrt{\q}\left[-6\sqrt{\p}\,({\k}^2+\frac{1}{r_o^2})\right]\nn\\
&=&-\frac{3V_o}{\kappa}\bar{N}\sqrt{\p}\,({\k}^2+\frac{1}{r_o^2}),\label{bhcc}
\ea
which naturally reproduces the background Hamiltonian constraint in the spatially flat case in the large $r_o$ limit.
\subsection{Perturbed constraints}
After including the perturbations, the canonical pair
 can be split as follows:
\ba
K_a^i&=&\bar{K}_a^i+\delta K_a^i=\k\,\,\wb^i_a+\delta K_a^i, \nn\\
E^a_i&=&\bar{E}^a_i+\delta E^a_i=\p\,\,\eb_i^a+\delta E^a_i.
\ea
Recall the definition of the spin connection,
\ba
\Gamma^i_a&\equiv&-\frac{1}{2}\epsilon^{ijk}E^b_j\big(\pa_aE^k_b-\pa_bE_a^k
+E_a^lE^c_k\pa_cE^l_b\nn\\
&&~+E^k_aE^d_m\pa_bE^m_d\big),\label{defGamma}
\ea
where $E^i_a$ denotes the inverse of $E^a_i$;
Direct calculation yields the variation of the spin connection,
\ba
\delta \Gamma^i_a&=&-\frac{1}{2}\epsilon^{ijk}E^b_j\big(D_a\delta E^k_b-D_b\delta E_a^k
+E_a^lE^c_kD_c\delta E^l_b\nn\\
&&~+E^k_aE^d_mD_b\delta E^m_d\big)\nn\\
&=&\frac{1}{2}\epsilon^{ijk}\big(E^k_bD_a\delta E^b_j-E^m_aE^k_dE^b_jD_b\delta
E^d_m\nn\\
&&~+E^l_aE^l_bE^c_kD_c\delta E^b_j+E^k_aE^m_dE^b_jD_b\delta E^d_m\big),\label{deltaGamma}
\ea
where $D$ denotes the covariant derivative with respect to $E^a_i$, note that in the second step we  used the identity $\delta E^k_a=-E^j_aE^k_b\delta E^b_j$. If we only consider the linear part, $\delta \Gamma^i_a$ can be approximated by
\ba
\delta\Gamma^i_a&\simeq&\frac{1}{2}\epsilon^{ijk}\big(\bar{E}^k_b\bar{D}_a\delta E^b_j-\bar{E}^m_a\bar{E}^k_d\bar{E}^b_j\bar{D}_b\delta E^d_m\nn\\
&&+\bar{E}^l_a\bar{E}^l_b\bar{E}^c_k\bar{D}_c\delta E^b_j+\bar{E}^k_a\bar{E}^m_d\bar{E}^b_j\bar{D}_b\delta E^d_m\big)\nn\\
&=&\frac{1}{2}\frac{1}{\bar{p}\,\sqrt{\q}}\big(\epsilon^{ijk}
\,\,\wb^k_b\,\delta^c_a
-\epsilon^{ilk}\,\,\eb^c_l\,\,\wb^k_b\,\,\wb^j_a\nn\\
&&+\epsilon^{ijk}\,\,\eb^c_k\,\,
\wb^l_a\,\,\wb^l_b+\epsilon^{ilk}\,\,\eb^c_l\,\,\wb^k_a\,\,\wb^j_b\big)
\bar{D}_c\delta E^b_j\nn\\
&\equiv&\frac{1}{2}\frac{1}{\bar{p}\,\sqrt{\q}}\,X^{ijc}_{ba}\,\bar{D}_c\delta E^b_j,\label{linearGamma}
\ea
where $\bar{D}$ denotes the derivative compatible with $\bar{E}^a_i$.

Using (\ref{linearGamma}), we derive the perturbed gravitational Hamiltonian constraint which consists of quadratic terms of linearly perturbed variables,
\ba
H_{g}[N]=\frac{1}{2 \kappa} \int_\mathcal{C} d^3x \left[\delta N \mathcal{H}^{(1)}_g+\bar{N}\mathcal{H}^{(2)}_g\right],
\ea
where
\ba
\mathcal{H}^{(1)}_g&=&
-4\sqrt{\q}\sqrt{\bar{p}}\,\bar{k}\,\,\eb^a_i\delta K^i_a
-\frac{1}{\sqrt{\bar{p}}}\,\,({\k}^2+\frac{1}{r_o^2})\,\,\wb^i_a\delta E^a_i\nn  \\
&&+\frac{2}{\sqrt{\bar{p}}}\,\, \eb^a_j\bar{D}_a \bar{D}_b \delta E^b_j ,  \\
\mathcal{H}^{(2)}_g &=&
\sqrt{\q}\sqrt{\bar{p}}\,\,\eb^a_i\,\eb^b_j\delta K^j_a \delta K^i_b
-\sqrt{\q}\sqrt{\bar{p}}\, (\,\eb^a_i\delta K^i_a )^2 \nn  \\
&&-\frac{2}{\sqrt{\bar{p}}}\,\bar{k} \,\delta E^a_i \delta K^i_a
+\frac{1}{4\sqrt{\q}\,\bar{p}^{\frac{3}{2}}}\,(\bar{k}^2+\frac{1}{r_o^2})\,( \,\wb^i_a\,\delta E^a_i)^2  \nn\\
&&-\frac{1}{2\sqrt{\q}\,\bar{p}^{\frac{3}{2}}}\,(\bar{k}^2+\frac{1}{r_o^2})
\,\,\wb^i_a\,\wb^j_b\,\delta E^a_j \delta E^b_i\nn\\
&&
+\frac{1}{\sqrt{\q}\,\bar{p}^{\frac{3}{2}}}W^{cdij}_{ab}\,( \bar{D}_c \delta E^a_i) (\bar{D}_d \delta E^b_j) ,
\ea
with $W^{cdij}$ defined by
\ba
W^{cdij}_{ab}=&&\frac{1}{2}\,\eb^e_{[m}\,\eb^f_{n]}X^{mjd}_{be}X^{nic}_{af}
-\epsilon^{im}{}_n\,\eb^e_mX^{njd}_{be}\delta^c_a\nn\\
&&-\epsilon^{li}{}_{k}\,\eb^c_l\,X^{kjd}_{ba}
+\frac{1}{2}\epsilon^{mn}{}_k\,\eb^c_m\,\,\eb^e_n\,\,\wb^i_a\,X^{kjd}_{be}.\label{defW}
\ea

 In addition, it is easy to verify that the perturbed gravitational diffeomorphism constraint on the spatially closed background can be expressed as
\ba
\textbf{D}_g[N^a] &=& \frac{1}{\kappa} \int_\mathcal{C} d^3x \delta N^a \big[ 2\bar{p}\sqrt{\q}\,\,\eb^b_i \bar{D}_{[a}\delta K^i_{b]}
-\bar{k}\,\,\wb^i_a \bar{D}_b \delta E^b_i \big],\nn\\
\label{diffconstr}
\ea
and the perturbed Gauss constraint reads
\begin{equation}
\textbf{G}[\Lambda^i] = \frac{1}{\kappa} \int_\mathcal{C} d^3x \delta\Lambda^i [\bar{p}\sqrt{\q}\,\epsilon_{ij}{}^k\,\eb^a_k\,\delta K^j_a
+\bar{k}\,\epsilon_{ij}{}^k\,\wb^j_a\,\delta E^a_k ].\\
\label{guassconstr}
\end{equation}

\section{anomaly-free constraints with inverse-volume corrections}
In the effective theory of LQC, the inverse powers of densitized triads appearing in the gravitational Hamiltonian constraint (\ref{mathH}) are subject to the ``inverse-volume corrections," which are usually assumed to depend only on the triads and can be described by
\ba
\langle\sqrt{|\det E|}\rangle\langle\frac{1}{\sqrt{|\det E|}}\rangle\rightarrow \alpha(E^a_i)
\ea
where the function $\alpha(E^a_i)\rightarrow 1$ in the classical limit. Since its detailed expression cannot be determined due to the quantization ambiguities, we will not specify the explicit form of $\alpha$ in this section.

\subsection{Inverse-volume corrections and constraints}
In LQG, neither the diffeomorphism constraint nor Gauss constraint is modified by quantum corrections, and only the Hamiltonian constraint is affected; therefore, in LQC, we only consider the quantum corrections in the Hamiltonian constraint, the
gravitational part with inverse-volume corrections of which can be split as
\ba
\textbf{H}^Q_g[N]&=&\frac{1}{2\kappa}\int_\mathcal{C} d^3x(\bar{N}+\delta N)(\bar{\alpha}+\alpha^{(1)}+\alpha^{(2)})\nn\\
&&\qquad\times[\mathcal{H}^{(0)}_g+\mathcal{\tilde{H}}^{(1)}_g
+\mathcal{\tilde{H}}^{(2)}_g],
\label{qHamcons}\ea
wherein $\bar{\alpha}$ represents the homogenous part of the inverse-volume corrections and $\alpha^{(1)}$ and $\alpha^{(2)}$ are separately the inhomogeneous parts to first and second orders. In the classical case, $\bar{\alpha}=1$, and both $\alpha^{(1)}$ and $\alpha^{(2)}$ vanish. For convenience of calculations, we parametrize them as
\ba
\bar{\alpha}&\equiv&\bar{\alpha}(\p),\\
\alpha^{(1)}&\equiv&\frac{\bar{\alpha}(\p)}{\sqrt{\q}\,\p}\,\alpha_1(\p)\,\,\wb_a^i\delta E^a_i,\\
\alpha^{(2)}&\equiv&\frac{\bar{\alpha}(\p)}{\q\,\p^2}\left[\alpha_2(\p)
\,\,\wb^i_a\,\wb^j_b\,\delta E^a_j \delta E^b_i
+\alpha_3(\p)\,( \,\wb^i_a\,\delta E^a_i)^2\right].\nn\\
\ea
The gravitational Hamiltonian densities $\mathcal{H}^{(0)}$, $\mathcal{\tilde{H}}^{(1)}$ and $\mathcal{\tilde{H}}^{(2)}$ separately read

\ba
\mathcal{H}^{(0)}_g&=&-6\sqrt{\q}\sqrt{\p}\,({\k}^2+\frac{1}{r_o^2}),\\
\mathcal{\tilde{H}}^{(1)}_g&=&-4\sqrt{\q}\sqrt{\bar{p}}\,\bar{k}\,(1+f_1)
\,\,\eb^a_i\delta K^i_a\nn\\
&&-\frac{1}{\sqrt{\bar{p}}}\,{\k}^2(1+f_2)\,\,\wb^i_a\delta E^a_i\nn\\
&&-\frac{1}{\sqrt{\bar{p}}}\,\frac{1}{r_o^2}(1+f_3)\,\,\wb^i_a\delta E^a_i\nn\\
&&+\frac{2}{\sqrt{\bar{p}}}(1+f_4)\, \,\eb^a_j\bar{D}_a \bar{D}_b \delta E^b_j ,\\
\mathcal{\tilde{H}}^{(2)}_g &=&
\sqrt{\q}\sqrt{\bar{p}}\,(1+f_5)\,\,\eb^a_i\,\eb^b_j\delta K^j_a \delta K^i_b\nn\\
&&-\sqrt{\q}\sqrt{\bar{p}}\,(1+f_6)\, (\,\eb^a_i\delta K^i_a )^2\nn\\ &&-\frac{2}{\sqrt{\bar{p}}}\,\bar{k}\,(1+f_7) \,\delta E^a_i \delta K^i_a\nn\\
&&+\frac{1}{4\sqrt{\q}\,\bar{p}^{\frac{3}{2}}}\,\bar{k}^2\,(1+f_8)\,( \,\wb^i_a\,\delta E^a_i)^2 \nn  \\
&&+\frac{1}{4\sqrt{\q}\,\bar{p}^{\frac{3}{2}}}\,\frac{1}{r_o^2}\,(1+f_9)\,( \,\wb^i_a\,\delta E^a_i)^2\nn\\
&&-\frac{1}{2\sqrt{\q}\,\bar{p}^{\frac{3}{2}}}\,\bar{k}^2\,(1+f_{10})
\,\,\wb^i_a\,\wb^j_b\,\delta E^a_j \delta E^b_i\nn\\
&&-\frac{1}{2\sqrt{\q}\,\bar{p}^{\frac{3}{2}}}\,\frac{1}{r_o^2}\,(1+f_{11})
\,\,\wb^i_a\,\wb^j_b\,\delta E^a_j \delta E^b_i\nn \\
&&+\frac{1}{\sqrt{\q}\,\bar{p}^{\frac{3}{2}}}\,(1+f_{12})\,W^{cdij}_{ab}\,( \bar{D}_c \delta E^a_i) (\bar{D}_d \delta E^b_j).\nn\\\label{qcgHam}
\ea
As mentioned above, the functions $f_1$ to $f_{12}$ are counterterms introduced to cancel the anomalies in the following algebra; in this paper, we assume that they depend only on the gravitational variable $\p$. From (\ref{qHamcons}) to (\ref{qcgHam}), it is easy to see that $\alpha_2(\p)$ and $\alpha_3(\p)$ can be absorbed into the counterterms in the second order expansion of the Hamiltonian constraint; therefore, without loss of generality, we set $\alpha_2=\alpha_3=0$. Since the counterterms cannot be uniquely fixed by the gravitational part alone and as shown in Ref. \cite{Bojowald:2008a}, the introduction of matter field can help  determine some counterterms; in this paper, we also consider the simple model of the minimally coupled scalar field with general potential shape $V(\varphi)$, and now the matter part of Hamiltonian constraint can be written as
\ba
\textbf{H}^Q_m[N]&=&\int_\mathcal{C} d^3x(\bar{N}+\delta N)\bigg[
\left(\bar{\beta}+\beta^{(1)}+\beta^{(2)}\right)\nn\\
&&\quad\times\left(\mathcal{H}^{(0)}_{\pi}
+\mathcal{\tilde{H}}^{(1)}_{\pi}+\mathcal{\tilde{H}}^{(2)}_{\pi}
+\mathcal{\tilde{H}}^{(2)}_{\nabla}\right)\nn\\
&&\quad+\left(\mathcal{H}^{(0)}_{\varphi}
+\mathcal{\tilde{H}}^{(1)}_{\varphi}+\mathcal{\tilde{H}}^{(2)}_{\varphi}\right)\bigg],
\ea
where similarly as the gravitational part, the inverse-volume corrections in the matter part are defined as
\ba
\bar{\beta}&\equiv&\bar{\beta}(\p),\\
\beta^{(1)}&\equiv&\frac{\bar{\beta}(\p)}{\sqrt{\q}\,\p}\,\beta_1(\p)\,\,\wb_a^i\delta E^a_i,\\
\beta^{(2)}&\equiv&\frac{\bar{\beta}(\p)}{\q\,\p^2}\left[\beta_2(\p)
\,\,\wb^i_a\,\wb^j_b\,\delta E^a_j \delta E^b_i
+\beta_3(\p)\,( \,\wb^i_a\,\delta E^a_i)^2\right],\nn\\
\ea
which arise due to the appearance of inverse densitized triads and the inverse metric in the kinetic sector, and must satisfy the requirement that $\bar{\beta}\rightarrow1$ in the classical limit. Generally speaking, they may take a different form from the correction $\alpha$.
The different parts in the matter Hamiltonian constraint read as follows,
\ba
\mathcal{H}^{(0)}_{\pi}&=&\frac{\bar{\pi}^{2}}{2\sqrt{\q}\,\p^{\frac{3}{2}}},\\
\mathcal{H}^{(0)}_{\varphi}&=&\sqrt{\q}\,\p^{\frac{3}{2}}V(\bar{\varphi}),\\
{\tilde{H}}^{(1)}_{\pi}&=&\frac{\bar{\pi}}{\sqrt{\q}\,\p^{\frac{3}{2}}}(1+g_1)\delta \pi\nn\\
&&-\frac{\bar{\pi}^{2}}{4\,\q\,\p^{\frac{5}{2}}}(1+g_2)\,\,\wb^i_a\delta E^a_i,\\
\ea
\ba
\mathcal{\tilde{H}}^{(1)}_{\varphi}&=&\sqrt{\q}
\,\p^{\frac{3}{2}}V_{,\varphi}(\bar{\varphi})(1+g_3)\delta \varphi\nn\\
&&+\frac{\sqrt{\p}}{2}V(\bar{\varphi})(1+g_4)\,\,\wb^i_a\delta E^a_i,\\
\mathcal{\tilde{H}}^{(2)}_{\pi}&=&\frac{1}{2\sqrt{\q}\,\p^{\frac{3}{2}}}(1+g_5)(\delta \pi)^2\nn\\
&&-\frac{\bar{\pi}}{2\,\q\,\p^{\frac{5}{2}}}(1+g_6)
\delta \pi\,\,\wb^i_a\delta E^a_i\nn\\
&&+\frac{\bar{\pi}^2}{8\,\q^{\frac{3}{2}}\p^{\frac{7}{2}}}
(1+g_7)\,\,\wb^i_a\,\wb^j_b\,\delta E^a_j \delta E^b_i\nn\\
&&+\frac{\bar{\pi}^2}{16\,\q^{\frac{3}{2}}\p^{\frac{7}{2}}}\,
(1+g_8)( \,\wb^i_a\,\delta E^a_i)^2,\\
\mathcal{\tilde{H}}^{(2)}_{\nabla}&=&\frac{\sqrt{\q}\,\sqrt{\p}}{2}(1+g_9)
\,\,\eb^a_i\,\,\eb^b_i
\,(\bar{D}_a\delta \varphi)\bar{D}_b\delta \varphi,
\ea
\ba
\mathcal{\tilde{H}}^{(2)}_{\varphi}&=&\frac{\sqrt{\q}\,\p^{\frac{3}{2}}}{2}
V_{,\varphi\varphi}(\bar{\varphi})(1+g_{10})\,(\delta \varphi)^2\nn\\
&&+\frac{\sqrt{\p}}{2}V_{,\varphi}(\bar{\varphi})(1+g_{11})
\,\delta \varphi\,\,\wb^i_a\delta E^a_i\nn\\
&&-\frac{V(\bar{\varphi})}{4\,\sqrt{\q}\,\sqrt{\p}}(1+g_{12})
\,\,\wb^i_a\,\wb^j_b\,\delta E^a_j \delta E^b_i\nn\\
&&+\frac{V(\bar{\varphi})}{8\,\sqrt{\q}\,\sqrt{\p}}(1+g_{13})\,(\,\wb^i_a\,\delta E^a_i)^2,\label{qcmHam}
\ea
where $V_{,\varphi}(\bar{\varphi})\equiv\frac{d V(\bar{\varphi})}{d \bar{\varphi}}$,\,$V_{,\varphi\varphi}(\bar{\varphi})\equiv\frac{d^2 V(\bar{\varphi})}{d \bar{\varphi}^2}$. Likewise,  the counterterms $g_1$ to $g_{13}$ are also assumed to be functions only of $\p$ and should vanish in the classical limit. Moreover, similarly as above, $\beta_1$, $\beta_2$ and $\beta_3$ can be absorbed into the counterterms; hence, in the following, we choose $\beta_1=\beta_{2}=\beta_3=0$.

The Hamiltonian constraint is given by
\ba
\textbf{H}^Q[N] &=& \textbf{H}^Q_g[N]+\textbf{H}^Q_m[N]
\ea
and the diffeomorphism constraint reads
\ba
\textbf{D}[N^a] &=& \textbf{D}_g[N^a]+\textbf{D}_m[N^a]
\ea
where the matter part $\textbf{D}_m[N^a]=\int_\mathcal{C} d^3x \,\delta N^a \bar{\pi}\bar{D}_a\delta \varphi$ retains the classical form just like the gravitational counterpart.
\subsection{Poisson brackets and anomalies}
In this section, we calculate the different Poisson brackets. Since the diffeomorphism constraint and Gauss constraint keep the classical form, the Poisson brackets between them do not generate anomalies; hence, the possible anomalies can only be generated from the brackets with quantum-corrected Hamiltonian constraint.
\subsubsection{$\{\textbf{H}^Q[N],\textbf{D}[N^a]\}$}
Using the elementary Poisson bracket
\ba
\{\delta K_a^i(x),\delta E^b_j(y)\}=\kappa\delta^i_j\delta_a^b\delta^3(x-y)
\ea
and the above expressions for $\textbf{H}_g^Q$ and $\textbf{H}_m^Q$, after direct calculations, we obtain the following result, which is written as sum of independent terms,
\ba
&&\{\textbf{H}^Q[N],\textbf{D}[N^a]\}\nn\\
&=&-\textbf{H}^Q\,[\delta N^a \bar{D}_a \delta N]\nn\\
&&+\frac{1}{\kappa}\int_\mathcal{C}d^3x \,\bar{\alpha}\,\sqrt{\q}\sqrt{\bar{p}}\,\k^2\,(\delta N^a \bar{D}_a \delta N)\,\mathcal{A}_1\nn\\
&&+\frac{1}{\kappa}\int_\mathcal{C}d^3x \,\bar{\alpha}\, \sqrt{\q}\sqrt{\bar{p}}\,\frac{1}{r_o^2}\,(\delta N^a \bar{D}_a \delta N)\,\mathcal{A}_2\nn\\
&&+\frac{1}{\kappa}\int_\mathcal{C} d^3x\,\bar{\alpha}\, \bar N\sqrt{\q}\sqrt{\bar{p}}\,\k\,(\delta N^c \,\eb^a_i\bar{D}_c \delta K^i_a)\,\mathcal{A}_3\nn\\
&&+\frac{1}{\kappa}\int_\mathcal{C} d^3x \,\,\bar{\alpha}\,\bar N\sqrt{\q}\sqrt{\bar{p}}\,\k\, (\delta N^a \,\eb^c_i \bar{D}_c \delta K^i_a)\,\mathcal{A}_4\nn\\
&&+\frac{1}{\kappa}\int_\mathcal{C}  d^3x\,\bar{\alpha}\, \frac{\bar N}{\sqrt{\bar{p}}}\,\k^2\,(\, \wb^i_a\delta N^c \bar{D}_c \delta E^a_i) \, \mathcal{A}_5\nn\\
&&+\frac{1}{\kappa}\int_\mathcal{C} d^3x\,\bar{\alpha}\, \frac{\bar N}{\sqrt{\bar{p}}}\,\k^2\,(\delta N^c\,\wb^i_c\bar{D}_a \delta E^a_i) \,\mathcal{A}_6\nn\\
&&+\frac{1}{\kappa}\int_\mathcal{C}  d^3x\,\bar{\alpha}\, \frac{\bar N}{\sqrt{\bar{p}}}\,\frac{1}{r_o^2}\,(\, \wb^i_a\delta N^c \bar{D}_c \delta E^a_i) \, \mathcal{A}_7\nn\\
&&+\frac{1}{\kappa}\int_\mathcal{C} d^3x\,\bar{\alpha}\, \frac{\bar N}{\sqrt{\bar{p}}}\,\frac{1}{r_o^2}\,(\delta N^c\,\wb^i_c\bar{D}_a \delta E^a_i) \,\mathcal{A}_8\nn\\
&&+\frac{1}{\kappa}\int_\mathcal{C} d^3x\, \bar N\sqrt{\bar{p}}\,(\bar{D}_a\delta N^a)(\,\eb^c_j\bar{D}_c \bar{D}_b \delta E^b_j )\,\mathcal{A}_9\nn\\
&&+\int_\mathcal{C} d^3x \,\sqrt{\q}\,\bar p^{\frac{3}{2}} V(\bar{\varphi}) \,(\delta N^c \bar{D}_c \delta N )\,\mathcal{A}_{10}\nn\\
&&+\int_\mathcal{C} d^3x\, \frac{\bar{\beta}\bar \pi^2}{2\sqrt{\q}\, \bar p^{\frac{3}{2}}}\,( \delta N^c \bar{D}_c \delta N)\, \mathcal{A}_{11}\nn\\
&&+\int_\mathcal{C} d^3x \,\bar N \sqrt{\q}\,\bar p^{\frac{3}{2}}V_{,\varphi}(\bar{\varphi})\,(\delta N^c\bar{D}_c \delta \varphi ) \, \mathcal{A}_{12}\nn\\
&&+\int_\mathcal{C} d^3x\, \frac{\bar{\beta}\bar N\bar \pi}{\sqrt{\q}\,\bar p^{\frac{3}{2}}}\, (\delta N^c\bar{D}_c \delta \pi ) \, \mathcal{A}_{13}\nn\\
&&+\int_\mathcal{C}  d^3x \,\bar N\frac{\sqrt{\bar{p}}\,V(\bar{\varphi})}{2}\, (\delta N^c\,\wb^i_c \bar{D}_a \delta E^a_i) \,\mathcal{A}_{14}\nn\\
&&+\int_\mathcal{C}  d^3x\,\bar N \frac{\sqrt{\bar{p}}\,V(\bar{\varphi})}{2} \,(\, \wb^i_a\delta N^c \bar{D}_c \delta E^a_i) \, \mathcal{A}_{15}\nn\\
&&+\int_\mathcal{C}  d^3x\, \frac{\bar{\beta}\bar N\bar{\pi}^2 }{12\,\q\,\bar{p}^{\frac{5}{2}}} \, (\delta N^c\,\wb^i_c \bar{D}_a \delta E^a_i) \,  \mathcal{A}_{16}\nn\\
&&+\int_\mathcal{C}  d^3x\, \frac{\bar{\beta}\bar N\bar{\pi}^2}{4\,\q\,\bar{p}^{\frac{5}{2}}} \,(\, \wb^i_a\delta N^c \bar{D}_c \delta E^a_i) \,\mathcal{A}_{17} ,
\ea
where $\mathcal{A}_{1}$ to $\mathcal{A}_{14}$ are anomalies which read explicitly as
\ba
\mathcal{A}_1&=&6\,\alpha_1+2f_1+f_2,\label{anomaly1}\\
\mathcal{A}_2&=&6\,\alpha_1+f_3+2f_4,\label{anomaly2}\\
\mathcal{A}_3&=&4(1+f_1)\,\alpha_1+ f_6+ f_7 ,\label{anomaly3}\\
\mathcal{A}_4&=&- f_5- f_7,\label{anomaly4}\\
\mathcal{A}_5&=&(4+2f_1+2f_2)\alpha_1+\frac{f_{10}}{2}-\frac{f_8}{2},\label{anomaly5}
\ea
\ba
\mathcal{A}_6&=&f_7-\frac{f_{10}}{2}-\frac{\bar{p}}{\bar{\alpha}} \frac{d \bar{\alpha} }{d \bar{p}},\label{anomaly6}\\
\mathcal{A}_7&=&(4+2f_3+2f_4)\,\alpha_1+\frac{f_{11}}{2}-\frac{f_9}{2},\label{anomaly7}\\
\mathcal{A}_8&=&-\frac{f_{11}}{2}+f_{12}-\frac{\bar{p}}{\bar{\alpha}} \frac{d \bar{\alpha} }{d \bar{p}},\label{anomaly8}\\
\mathcal{A}_9&=&2\alpha_1(1+f_4),\label{anomaly9}\\
\mathcal{A}_{10}&=&-g_4,\label{anomaly10}\\
\mathcal{A}_{11}&=&g_2-2g_1,\label{anomaly11}\\
\mathcal{A}_{12}&=&-g_{11},\label{anomaly12}\\
\mathcal{A}_{13}&=&g_6-g_5,\label{anomaly13}\\
\mathcal{A}_{14}&=&-g_{12},\label{anomaly14}\\
\mathcal{A}_{15}&=&g_{12}-g_{13},\label{anomaly15}\\
\mathcal{A}_{16}&=&3g_7+2\frac{\bar{p}}{\bar{\beta}} \frac{d \bar{\beta} }{d \bar{p}},\label{anomaly16}\\
\mathcal{A}_{17}&=&2g_6-g_8-g_7.\label{anomaly17}
\ea
Note that in deriving the anomalies $\mathcal{A}_2$ and $\mathcal{A}_8$ we have used the identity
\ba
&&\left\{\int_\mathcal{C} d^3x\frac{\delta N}{\sqrt{\p}}\,\, \eb^a_j\bar{D}_a \bar{D}_b \delta E^b_j ,\quad\int_\mathcal{C} d^3x\delta N^a\bar{p}\sqrt{\q}\,\,\eb^b_i \bar{D}_{[a}\delta K^i_{b]}\right\}\nn\\
&&=\kappa\int_\mathcal{C}d^3x \sqrt{\q}\sqrt{\bar{p}}\,\,\frac{1}{r_o^2}\,\delta N^a \bar{D}_a \delta N,\label{formula1}
\ea
to which we will give a proof in the Appendix.

\subsubsection{$\{\textbf{H}^Q[N_1],\textbf{H}^Q[N_2]\}$}
The Poisson brackets between the Hamiltonian constraints lead to
\ba
&&\{\textbf{H}^Q[N_1],\textbf{H}^Q[N_2]\}\nn\\
&&=\bar{\alpha}^2(1+f_4)(1+f_6)\textbf{D} \left[ \frac{\bar{N}}{\bar{p}} \bar{D}^a(\delta N_2 -\delta N_1)  \right]\nn\\
&&+\frac{1}{\kappa}\int_\mathcal{C} d^3x \,\bar{\alpha}^2\bar{N}\sqrt{\q}\,\bar{D}^a(\delta N_2 -\delta N_1)(\,\eb^b_i\bar{D}_b \delta K^i_a)\mathcal{A}_{18}\nn\\
&&+\frac{1}{\kappa}\int_\mathcal{C} d^3x\,\bar{\alpha}^2\bar{N}\frac{\k}{\bar{p}} \bar{D}^a(\delta N_2 -\delta N_1)(\,\wb^i_a\bar{D}_c \delta E^c_i) \mathcal{A}_{19}\nn\\
&&+\frac{1}{\kappa}\int_\mathcal{C} d^3x\,\bar{\alpha}^2\bar{N}\frac{\k}{\bar{p}} \bar{D}^c(\delta N_2 -\delta N_1)(\,\wb^i_a\bar{D}_c \delta E^a_i) \mathcal{A}_{20}\nn\\
&&+\frac{1}{\kappa}\int_\mathcal{C} d^3x\,\bar{\alpha}^2\bar{N}\sqrt{\q}\,\k^2 (\delta N_2 -\delta N_1)(\,\eb^a_i  \delta K^i_a) \mathcal{A}_{21}\nn\\
&&+\frac{1}{\kappa}\int_\mathcal{C} d^3x\,\bar{\alpha}^2\bar{N}\sqrt{\q}\,\frac{1}{r_o^2}\, (\delta N_2 -\delta N_1)(\,\eb^a_i  \delta K^i_a) \mathcal{A}_{22}\nn\\
&&+\frac{1}{\kappa}\int_\mathcal{C} d^3x\,\bar{\alpha}^2\frac{\bar{N}}{\p}\,\k^3 (\delta N_2 -\delta N_1)(\,\wb_a^i  \delta E_i^a) \mathcal{A}_{23}\nn\\
&&+\frac{1}{\kappa}\int_\mathcal{C} d^3x\,\bar{\alpha}^2\frac{\bar{N}}{\p}\,\frac{\k}{r_o^2}\,(\delta N_2 -\delta N_1)(\,\wb_a^i  \delta E_i^a) \mathcal{A}_{24}\nn
\ea
\ba
&&+\int_\mathcal{C}d^3x \,\bar{\alpha}\bar{\beta}\frac{\bar{N}}{\sqrt{\q}\bar{p}^2}\frac{\bar{\pi}^2}{2} (\delta N_2-\delta N_1)(\,\eb^a_i\delta K^i_a) \mathcal{A}_{25}\nn\\
&&+\int_\mathcal{C}d^3x\,\bar{\alpha} \bar{N}\sqrt{\q}\bar{p}\,V(\bar{\varphi})\,(\delta N_2-\delta N_1)
(\,\eb^a_i\delta K^i_a)  \mathcal{A}_{26}\nn\\
&&+\int_\mathcal{C}d^3x \,\bar{\alpha}\bar{\beta}\frac{\bar{N}}{\q\bar{p}^3}\frac{ \bar{\pi}^2 }{4}(\delta N_2-\delta N_1)
(\,\wb_a^i\delta E^a_i)\mathcal{A}_{27}\nn\\
&&+\int_\mathcal{C}d^3x\,\bar{\alpha} \bar{N}\frac{V(\bar{\varphi})}{2}(\delta N_2-\delta N_1)
(\,\wb_a^i\delta E^a_i)  \mathcal{A}_{28}\nn\\
&&+\int_\mathcal{C}d^3x \,\bar{\alpha}\bar{\beta}\frac{\bar{N}}{\sqrt{\q}\bar{p}^2}\bar{\pi}(\delta N_2-\delta N_1)\,\delta \pi\mathcal{A}_{29}\nn\\
&&+\int_\mathcal{C}d^3x \,\bar{\alpha}\bar{N}\sqrt{\q}\bar{p}\,V_{,\varphi}(\bar{\varphi})\,(\delta N_2-\delta N_1)\,\delta \varphi\mathcal{A}_{30}\nn\\
&&+\int_\mathcal{C}d^3x \,\bar{\beta}\frac{\bar{N}}{\sqrt{\q}\bar{p}}\frac{\bar{\pi} V_{,\varphi}(\bar{\varphi})}{2}(\delta N_2-\delta N_1)(\,\wb_a^i\delta E^a_i) \mathcal{A}_{31}\nn\\
&&+\int_\mathcal{C}d^3x\,\bar{\beta} \bar{N}\,\bar{\pi} V_{,\varphi\varphi}(\bar{\varphi})(\delta N_2-\delta N_1)
(\delta \varphi)\mathcal{A}_{32}\nn\\
&&+\int_\mathcal{C}d^3x\,\bar{\beta} \bar{N}V_{,\varphi}(\bar{\varphi})(\delta N_2-\delta N_1)
\,\delta \pi\mathcal{A}_{33} \nn\\
&&+\int_\mathcal{C} d^3x\,\bar{\beta}^2\frac{\bar{N}}{\bar{p}}\bar{\pi} \bar{D}^a(\delta N_2 -\delta N_1)(\bar{D}_a\delta\varphi) \mathcal{A}_{34},
\ea
where $\bar{D}^a\equiv\,\eb^a_i\,\eb^b_i\bar{D}_b$ and the anomalies are given by
\ba
\mathcal{A}_{18}&=&(1+f_4)(f_6 - f_5) ,\label{anomaly18}\\
\mathcal{A}_{19}&=&(1+f_4)(1+f_6+f_7+2\frac{\bar{p}}{\bar{\alpha}}\frac{d \bar{\alpha}}{d \bar{p}})+2\bar{p}\frac{d f_4}{d \bar{p}}
 \nn  \\
 &&-(1+f_1)(1+f_{12}+6\alpha_1) ,\label{anomaly19}\\
 \mathcal{A}_{20}&=&2(f_4-f_1)\,\alpha_1,\label{anomaly20}\\
\mathcal{A}_{21}&=&(1+f_1)(-1+2\frac{\bar{p}}{\bar{\alpha}}\frac{d \bar{\alpha}}{d \bar{p}}-2f_7-12\alpha_1-12\alpha_1f_1)\nonumber  \\
 &&+4\bar{p}\frac{d f_1}{d \bar{p}}+\frac{1}{2}(1+f_2+6\alpha_1)(2+3f_6-f_5),\label{anomaly21}\\
\mathcal{A}_{22}&=&-(1+2\frac{\bar{p}}{\bar{\alpha}}\frac{d \bar{\alpha}}{d \bar{p}})(1+f_1)+\frac{1}{2}(1+f_3+6\alpha_1)\nn\\
&&\times(2+3f_6-f_5),\label{anomaly22}\\
\mathcal{A}_{23}&=&\frac{1}{2}(-1+f_7+6\alpha_1+6\alpha_1f_1)(1+f_2+6\alpha_1)\nn\\
 &&+\frac{1}{2}(1+f_1)(1+3f_8-12\alpha_1-12\alpha_1f_2-2f_{10}) ,\nn\\
 &&+\bar{p}\frac{d f_2}{d \bar{p}}+6\bar{p}\frac{d \alpha_1}{d \bar{p}}, \label{anomaly23}\\
\mathcal{A}_{24}&=&(\frac{1}{2}f_7+\frac{\bar{p}}{\bar{\alpha}}\frac{d \bar{\alpha}}{d \bar{p}}+3\alpha_1+3\alpha_1f_1)(1+f_3+6\alpha_1)\nn\\
 &&+\frac{1}{2}(1+f_1)(1+3f_9-2f_{11}-12\alpha_1+3\alpha_1f_3)\nn\\
 && -\frac{1}{2}(1+2\frac{\bar{p}}{\bar{\alpha}}\frac{d \bar{\alpha}}{d \bar{p}})(1 +f_2+6\alpha_1)+\bar{p}\frac{d f_3}{d \bar{p}} \nn\\
&&+6\bar{p}\frac{d \alpha_1}{d \bar{p}},\label{anomaly24}
\ea
\ba
\mathcal{A}_{25}&=&(1+g_2)(1+\frac{3}{2}f_6-\frac{1}{2}f_5)
 -(1-\frac{2}{3}\frac{\p}{\bar{\beta}}\frac{d\bar{\beta}}{d\p})\nn\\
 &&\times(1+f_1),\label{anomaly25}\\
\mathcal{A}_{26}&=&(1+f_1)-(1+g_4)(1+\frac{3}{2}f_6-\frac{1}{2}f_5) ,\label{anomaly26}\\
\mathcal{A}_{27}&=&(1+g_2)(1+f_7+6\alpha_1+6\alpha_1f_1)\nn\\
 &&+(\frac{2}{3}\frac{\p}{\bar{\beta}}\frac{d\bar{\beta}}{d\p}-1)
 (1+f_2+6\alpha_1)\nn\\
 &&+(-5+2\frac{\p}{\bar{\beta}}\frac{d\bar{\beta}}{d\p})(1+g_2)+2\p\frac{d g_2}{d\p},\nn\\
 &&+(1+f_1)(5+3g_8+2g_7),\label{anomaly27}\\
\mathcal{A}_{28}&=&f_2+6\alpha_1-(1+f_7+6\alpha_1+6\alpha_1f_1)(1+g_4)\nn\\
 &&+(1+f_1)(1+3g_{13}-2g_{12})-g_4-2\p\frac{d g_4}{d\p},\label{anomaly28}\\
\mathcal{A}_{29}&=&-3(1+f_1)(1+g_6)+(1+g_1)
 (3-2\frac{\p}{\bar{\beta}}\frac{d\bar{\beta}}{d\p})\nn\\
 &&-2\p\frac{d g_1}{d\p},\label{anomaly29}\\
\mathcal{A}_{30}&=&3(1+f_1)(1+g_{11})-3(1+g_3)-2p\frac{d g_3}{d \bar{p}},\label{anomaly30}
\ea
\ba
\mathcal{A}_{31}&=&-2-g_2-g_4+(1+g_3)(1+g_6)\nn\\
 &&+(1+g_1)(1+g_{11}) ,\label{anomaly31}\\
\mathcal{A}_{32}&=&-(1+g_3)+(1+g_1)(1+g_{10}) ,\label{anomaly32}\\
\mathcal{A}_{33}&=&1+g_1-(1+g_3)(1+g_5) ,\label{anomaly33}\\
\mathcal{A}_{34}&=&\bar{\beta}^2(1+g_1)(1+g_9)-\bar{\alpha}^2(1+f_4)(1+f_6).
\label{anomaly34}
 \ea

Notice that to derive equation (\ref{anomaly19}) we used the bracket
\ba
&&\bigg\{\int_\mathcal{C} d^3x \,\delta N\sqrt{\q}(\,\eb^a_i\delta K^i_a),\int_\mathcal{C} d^3x \,\frac{\bar{N}}{\sqrt{\q}}\,W^{cdij}_{ab}( \bar{D}_c \delta E^a_i) \nn\\
&&\quad\times(\bar{D}_d \delta E^b_j) \bigg\}=\kappa\int_\mathcal{C} d^3x \bar{N}\delta N (\,\eb^a_i\bar{D}_a\bar{D}_c \delta E^c_i) ,\label{bracketZk}
\ea
which can be verified by straight calculation using the definition of $W^{cdij}_{ab}$ in (\ref{defW}).

\subsubsection{$\{\textbf{H}^Q[N],\textbf{G}[\Lambda^i]\}$}
Finally, this Poisson bracket is given by
\ba
&&\{\textbf{H}^Q[N],\textbf{G}[\Lambda^i]\}\nn\\
&&=\frac{1}{\kappa}\int_\mathcal{C} d^3x \,\bar{\alpha} \bar{N}\sqrt{\q}\sqrt{\bar{p}}\,\k\,\epsilon_{ij}{}^{k}\,\eb^a_k\,\delta \Lambda^i\delta K_a^j\mathcal{A}_{35}\nn\\
&&+\frac{1}{\kappa}\int_\mathcal{C} d^3x\,\bar{\alpha}\frac{\bar{N}}{\sqrt{\bar{p}}}\k^2\epsilon_{ij}^{~~k}\,\wb^j_a\,\delta \Lambda^i\delta E^a_k\mathcal{A}_{36}\nn\\
&&+\frac{1}{\kappa}\int_\mathcal{C} d^3x\,\bar{\alpha}\frac{\bar{N}}{\sqrt{\bar{p}}}\,\frac{1}{r_o^2}
\,\epsilon_{ij}^{~~k}\,\wb^j_a\,\delta \Lambda^i\delta E^a_k\mathcal{A}_{37}\nn\\
&&+\int_\mathcal{C}  d^3x\, \frac{\bar{\beta}\bar N\bar{\pi}^2 }{12\,\q\,\bar{p}^{\frac{5}{2}}}\epsilon_{ij}^{~~k}\,\wb^j_a\,\delta \Lambda^i\delta E^a_k\mathcal{A}_{38}\nn\\
&&+\int_\mathcal{C}  d^3x \,\bar N\frac{\sqrt{\bar{p}}\,V(\bar{\varphi})}{2}\epsilon_{ij}^{~~k}\,\wb^j_a\,\delta \Lambda^i\delta E^a_k\mathcal{A}_{39},\label{anomaly39}
\ea
where
\ba
\mathcal{A}_{35}&=&f_5+f_7,\label{anomaly35}\\
\mathcal{A}_{36}&=&\frac{f_{10}}{2}-f_7+\frac{\bar{p}}{\bar{\alpha}}\frac{d \bar{\alpha}}{d \bar{p}},\label{anomaly36}\\
\mathcal{A}_{37}&=&\frac{f_{11}}{2}-f_{12}+\frac{\bar{p}}{\bar{\alpha}}\frac{d \bar{\alpha}}{d \bar{p}},\label{anomaly37}\\
\mathcal{A}_{38}&=&-3g_7-2\frac{\bar{p}}{\bar{\beta}} \frac{d \bar{\beta} }{d \bar{p}},\label{anomaly38}\\
\mathcal{A}_{39}&=&g_{12}.\label{anomaly39}
\ea

$\mathcal{A}_{1}$ to $\mathcal{A}_{38}$ are anomalies that should vanish on the effective level; i.e. we have to solve the equations
\ba
\mathcal{A}_i=0 ,\quad(i=1,2...39).\nonumber
\ea
\subsection{Solutions of the anomalies}
  Notice that in the anomaly equations there exist three unspecified inverse-volume correction functions $\bar{\alpha}$, $\alpha_1$ and $\bar{\beta}$ and 25 counterterms, which are $f_i$ $(i=1,2...12)$ and $g_i$ $(i=1,2...13)$, i.e. totally 28 unknown quantities. As shown above, to fix these quantities we need to solve up to 39 anomaly equations. However, not all of these equations are independent of each other; for instance, it is obvious that Eqs. (\ref{anomaly35}),(\ref{anomaly36}),(\ref{anomaly37}),(\ref{anomaly38}) and (\ref{anomaly39}) are separately equivalent to Eqs. (\ref{anomaly4}),(\ref{anomaly6}),(\ref{anomaly8}),(\ref{anomaly16}) and (\ref{anomaly14}), moreover, some other complex equations can be deduced from other ones, and thus the number of independent anomalies is less than the total. In the following we will give a sketch of the derivation for the counterterms.

 Let us begin with the simple ones. From Eqs. (\ref{anomaly9}),(\ref{anomaly10}),(\ref{anomaly12}),(\ref{anomaly14}) and
(\ref{anomaly15}), it is easy to see
\ba
\alpha_1=g_4=g_{11}=g_{12}=g_{13}=0.
\ea
Substituting it into (\ref{anomaly3}) and (\ref{anomaly26}),   using (\ref{anomaly1}),  (\ref{anomaly4}), (\ref{anomaly11}), (\ref{anomaly16}) and (\ref{anomaly25}), we have
\ba
f_1=-\frac{f_2}{2}=f_5=f_6=-f_7,\label{f1}
\ea
and
\ba
g_2=2g_1=g_7=-\frac{2}{3}\frac{\p}{\bar{\beta}}\frac{d\bar{\beta}}{d\p}.\label{g2}
\ea
Using (\ref{g2}) and (\ref{anomaly13}), then solving (\ref{anomaly29}), (\ref{anomaly30}) and (\ref{anomaly33}) simultaneously, we get
\ba
g_6&=&g_5=\bar{\beta}\left(1-\frac{1}{3}\frac{\p}{\bar{\beta}}\frac{d\bar{\beta}}{d\p}
\right)^2-1,\\
g_3&=&\frac{1}{\bar{\beta}\left(1-\frac{1}{3}
\frac{\p}{\bar{\beta}}\frac{d\bar{\beta}}{d\p}\right)}-1,\label{g3}\\
f_1&=&\frac{1-\frac{7}{9}\frac{\p}{\bar{\beta}}\frac{d\bar{\beta}}{d\p}
+\frac{2}{9}\frac{\p^2}{\bar{\beta}}
\frac{d^2\bar{\beta}}{d\p^2}}{\bar{\beta}\left(1-\frac{1}{3}\frac{\p}{\bar{\beta}}
\frac{d\bar{\beta}}{d\p}\right)^{2}}-1.\label{expressionf1}
\ea
Inserting (\ref{f1}) into (\ref{anomaly5}), (\ref{anomaly6}) and (\ref{anomaly22}), it is easy to see
\ba
f_{10}&=&-2f_1-2\frac{\bar{p}}{\bar{\alpha}}\frac{d \bar{\alpha}}{d \bar{p}},\\
f_8&=&f_{10},\\
f_3&=&2\frac{\bar{p}}{\bar{\alpha}}\frac{d \bar{\alpha}}{d \bar{p}};\label{f3}
\ea
substituting (\ref{f3}) into (\ref{anomaly2}), then using (\ref{f1}), (\ref{anomaly3}) and (\ref{anomaly19}), we have
\ba
f_4&=&-\frac{f_3}{2}=-\frac{\bar{p}}{\bar{\alpha}}\frac{d \bar{\alpha}}{d \bar{p}},\\
f_{12}&=&\frac{(1+f_4)(1+2\frac{\bar{p}}{\bar{\alpha}}\frac{d \bar{\alpha}}{d \bar{p}})+2\bar{p}\frac{d f_4}{d \bar{p}}}{1+f_1}-1,
\ea
and then from (\ref{anomaly7}) and (\ref{anomaly8}), we derive
\ba
f_{11}&=&2f_{12}-2\frac{\bar{p}}{\bar{\alpha}}\frac{d \bar{\alpha}}{d \bar{p}},\\
f_9&=&f_{11}.
\ea
Furthermore, using (\ref{f1}), (\ref{g2}) and (\ref{g3}) from (\ref{anomaly17}), (\ref{anomaly32}) and (\ref{anomaly34}), we get
\ba
g_8&=&2g_6+\frac{2}{3}\frac{\p}{\bar{\beta}}\frac{d\bar{\beta}}{d\p},\\
g_9&=&\frac{\bar{\alpha}^2}{\bar{\beta}^2}\frac{(1-\frac{\bar{p}}{\bar{\alpha}}\frac{d \bar{\alpha}}{d \bar{p}})}{(1-\frac{1}{3}\frac{\p}{\bar{\beta}}\frac{d\bar{\beta}}{d\p})}
(1+f_1)-1,\\
g_{10}&=&\frac{1}{\bar{\beta}\left(1-\frac{1}{3}\frac{\p}{\bar{\beta}}
\frac{d\bar{\beta}}{d\p}
\right)^2}-1.
\ea

  Note that we have derived the solutions of all counterterms, yet we still need to check whether these solutions satisfy the other anomaly equations which have not been used in the derivation. First, it is not difficult to check the above solutions solve the anomalies $\mathcal{A}_{18}$,~$\mathcal{A}_{20}$,
  ~\,$\mathcal{A}_{27}$,~$\mathcal{A}_{31}$, and next, we substitute the solutions into the remaining two equations (\ref{anomaly21}) and (\ref{anomaly23}); from both, we get the same equation:
\ba
2\p\frac{d f_1}{d\p}+(1+f_1)\frac{\bar{p}}{\bar{\alpha}}\frac{d \bar{\alpha}}{d \bar{p}}=0.\label{cof1}
\ea
Combined with the condition $\alpha(\p)\rightarrow 1$ in the classical limit, Eq. (\ref{cof1}) yields
\ba
f_1=\frac{1}{\sqrt{\bar{\alpha}}}-1;\label{solf1}
\ea
plugging (\ref{solf1}) into (\ref{expressionf1}), we get the equation
\ba
\frac{9-7\frac{\p}{\bar{\beta}}\frac{d\bar{\beta}}{d\p}
+2\frac{\p^2}{\bar{\beta}}
\frac{d^2\bar{\beta}}{d\p^2}}{\bar{\beta}\left(3-\frac{\p}{\bar{\beta}}
\frac{d\bar{\beta}}{d\p}\right)^{2}}=\frac{1}{\sqrt{\bar{\alpha}}},\label{consiscon}
\ea
which relates $\beta$ to $\alpha$ and it means that in order to get a consistent algebra the choices of inverse-volume corrections for the gravitational part should be relevant to the matter part.

Finally, using (\ref{consiscon}), we rewrite the counterterms as
\ba
f_1&=&\frac{1}{\sqrt{\bar{\alpha}}}-1,\\
f_2&=&-2(\frac{1}{\sqrt{\bar{\alpha}}}-1),\\
f_3&=&2\frac{\bar{p}}{\bar{\alpha}}\frac{d \bar{\alpha}}{d \bar{p}},\\
f_4&=&-\frac{\bar{p}}{\bar{\alpha}}\frac{d \bar{\alpha}}{d \bar{p}},\\
f_5&=&\frac{1}{\sqrt{\bar{\alpha}}}-1,\\
f_6&=&\frac{1}{\sqrt{\bar{\alpha}}}-1,\\
f_7&=&-(\frac{1}{\sqrt{\bar{\alpha}}}-1),\\
f_8&=&-\frac{2}{\sqrt{\bar{\alpha}}}+2-2\frac{\bar{p}}{\bar{\alpha}}\frac{d \bar{\alpha}}{d \bar{p}},\\
f_9&=&2\left(\sqrt{\bar{\alpha}}-1-\frac{\bar{p}}{\sqrt{\bar{\alpha}}}\frac{d \bar{\alpha}}{d \bar{p}}-\frac{\bar{p}}{\bar{\alpha}}\frac{d \bar{\alpha}}{d \bar{p}}-\frac{2\bar{p}^2}{\sqrt{\bar{\alpha}}}\frac{d^2 \bar{\alpha}}{d \bar{p}^2}\right),\\
f_{10}&=&-\frac{2}{\sqrt{\bar{\alpha}}}+2-2\frac{\bar{p}}{\bar{\alpha}}\frac{d \bar{\alpha}}{d \bar{p}},\\
f_{11}&=&2\left(\sqrt{\bar{\alpha}}-1-\frac{\bar{p}}{\sqrt{\bar{\alpha}}}\frac{d \bar{\alpha}}{d \bar{p}}-\frac{\bar{p}}{\bar{\alpha}}\frac{d \bar{\alpha}}{d \bar{p}}-\frac{2\bar{p}^2}{\sqrt{\bar{\alpha}}}\frac{d^2 \bar{\alpha}}{d \bar{p}^2}\right),\nn\\~\\
f_{12}&=&\sqrt{\bar{\alpha}}\left(1-\frac{\bar{p}}{\bar{\alpha}}\frac{d \bar{\alpha}}{d \bar{p}}-2\frac{\bar{p}^2}{\bar{\alpha}}\frac{d^2 \bar{\alpha}}{d \bar{p}^2}\right)-1,\\
g_1&=&-\frac{1}{3}\frac{\p}{\bar{\beta}}\frac{d\bar{\beta}}{d\p},\\
g_2&=&-\frac{2}{3}\frac{\p}{\bar{\beta}}\frac{d\bar{\beta}}{d\p},\\
g_3&=&\frac{1}{\bar{\beta}\left(1-\frac{1}{3}\frac{\p}{\bar{\beta}}
\frac{d\bar{\beta}}{d\p}\right)}-1,\\
g_4&=&0,\\
g_5&=&\bar{\beta}\left(1-\frac{1}{3}\frac{\p}{\bar{\beta}}\frac{d\bar{\beta}}{d\p}
\right)^2-1,\\
g_6&=&\bar{\beta}\left(1-\frac{1}{3}\frac{\p}{\bar{\beta}}\frac{d\bar{\beta}}{d\p}
\right)^2-1,\\
g_7&=&-\frac{2}{3}\frac{\p}{\bar{\beta}}\frac{d\bar{\beta}}{d\p},\\
g_8&=&2\left(\bar{\beta}\left(1-\frac{1}{3}\frac{\p}{\bar{\beta}}\frac{d\bar{\beta}}{d\p}
\right)^2+\frac{1}{3}\frac{\p}{\bar{\beta}}\frac{d\bar{\beta}}{d\p}-1\right),\\
g_9&=&\frac{\bar{\alpha}^\frac{3}{2}}{\bar{\beta}^2}
\frac{(1-\frac{\bar{p}}{\bar{\alpha}}
\frac{d\bar{\alpha}}{d\bar{p}})}{(1-\frac{1}{3}\frac{\p}{\bar{\beta}}
\frac{d\bar{\beta}}{d\p})}-1,\\
g_{10}&=&\frac{1}{\bar{\beta}\left(1-\frac{1}{3}
\frac{\p}{\bar{\beta}}\frac{d\bar{\beta}}{d\p}\right)^2}-1,\\
g_{11}&=&g_{12}=g_{13}=0,
\ea

 Now every counterterm has been uniquely expressed in terms of the background quantum corrections $\bar{\alpha}(\p)$ and $\bar{\beta}(\p)$, which indicates a consistent constraint algebra  can also be obtained in the spatially closed case.

  The nonvanishing Poisson brackets read
\ba
&&\{\textbf{H}^Q[N],\textbf{D}[N^a]\}=-\textbf{H}^Q\,[\delta N^a \bar{D}_a \delta N],\\
&&\{\textbf{H}^Q[N_1],\textbf{H}^Q[N_2]\}\nn\\
&&=\bar{\alpha}^\frac{3}{2}
(1-\frac{\bar{p}}{\bar{\alpha}}\frac{d\bar{\alpha}}{d\bar{p}})\textbf{D}
\left[ \frac{\bar{N}}{\bar{p}} \bar{D}^a(\delta N_2 -\delta N_1) \right].\label{brakHH}
\ea
The prefactor on the right-hand of (\ref{brakHH}) clearly shows that the algebra between the Hamiltonian constraints is deformed by the inverse-volume corrections.

Remarkably, the expressions of counterterms we have obtained on the spatially closed background are in complete agreement with the ones derived on the spatially flat background  \cite{Cailleteau:2013}
, which could be regarded as a cross-check of the correctness of our results and it also implies a uniform deformed algebra could exist for different spatial topology.
\subsection{Anomaly-free constraints on spatially flat FRW background}
 To recover the anomaly-free Hamiltonian constraint on the spatially flat background, we simply discard the terms proportional to the spatial curvature by setting $r_o\rightarrow\infty$ and then replace the derivative $\bar{D}_a$ with $\partial_a$. Furthermore, $\wb^i_a$, and $\eb^a_i$ are also replaced by $\delta^i_a$ and $\delta^a_i$, and $\sqrt{\q}$ is dropped to agree with the notations used in the spatially flat case \cite{Bojowald:2008a,Cailleteau:2011a}. For clarity,  in the following, we define
\ba
\Lambda_{\alpha}&\equiv&-\frac{\bar{p}}{\bar{\alpha}}\frac{d\bar{\alpha}}{d\bar{p}},\quad
\Omega_{\alpha}\equiv\frac{\bar{p}^2}{\bar{\alpha}}\frac{d^2\bar{\alpha}}{d\bar{p}^2},
\quad\Lambda_{\beta}\equiv-\frac{1}{3}\frac{\p}{\bar{\beta}}\frac{d\bar{\beta}}{d\p},\nn\\
\ea
which all vanish in the classical limit.

Using the counterterms derived above, the inverse-volume-corrected Hamiltonian constraint on the flat FRW background is expressed as
\ba
\textbf{H}^Q_{k=0}[N]=\textbf{H}_{k=0}^B[N]+\textbf{H}_{k=0}^P[N],
\ea
where $\textbf{H}_{k=0}^B[N]$ denotes the background part
\ba
\textbf{H}_{k=0}^B[N]=\int_\mathcal{C} d^3x\bar{N}\left[-\frac{3}{\kappa}\bar{\alpha}\sqrt{\p}\,{\k}^2
+\frac{\bar{\beta}\bar{\pi}^{2}}{2\,\p^{\frac{3}{2}}}
+\p^{\frac{3}{2}}V(\bar{\varphi})\right],\nn\\
\ea
and $\textbf{H}_{k=0}^P[N]$ denotes the perturbed part
\ba
\textbf{H}_{k=0}^P[N]=\textbf{H}_g^P[N]+\textbf{H}_m^P[N],
\ea
with
\ba
\textbf{H}^{P}_g[N] &=&\frac{1}{2 \kappa} \int_\mathcal{C} d^3x \left[\delta N \mathcal{H}^{(1)}_g+\bar{N}\mathcal{H}^{(2)}_g\right],\\
\textbf{H}^{P}_m[N] &=&\int_\mathcal{C} d^3x \left[\delta N \mathcal{H}^{(1)}_m+\bar{N}\mathcal{H}^{(2)}_m\right],
\ea
where

\ba
\mathcal{H}^{(1)}_g&=&
-4\,\sqrt{\bar{\alpha}}\sqrt{\bar{p}}\,\bar{k}\,\delta^a_i\delta K^i_a
-(3\bar{\alpha}-2\sqrt{\bar{\alpha}})\frac{\bar{k}^2}{\sqrt{\bar{p}}}
\,\delta^i_a\delta E^a_i\nn\\
&+&\bar{\alpha}\,(1+\Lambda_\alpha)\frac{2}{\sqrt{\bar{p}}}\,\partial_a \partial^i \delta E^a_i ,  \label{H1g}\\
\mathcal{H}^{(2)}_g &=&
\,\sqrt{\bar{\alpha}}\sqrt{\bar{p}}\,\delta^a_k \delta^b_j\delta K^j_a \delta K^k_b
-\,\sqrt{\bar{\alpha}}\sqrt{\bar{p}}\,(\delta^a_i\delta K^i_a )^2 \nn\\
&-&(2\bar{\alpha}-\sqrt{\bar{\alpha}})\frac{2\bar{k}}{\sqrt{\bar{p}}}\, \delta E^a_i \delta K^i_a\nn\\
&-&(3\bar{\alpha}-2\sqrt{\bar{\alpha}}+2\,\bar{\alpha}\Lambda_\alpha)
\,\frac{\bar{k}^2}{2\bar{p}^{\frac{3}{2}}}
\,\delta E^a_j \delta E^b_k \delta^k_a\delta^j_b \nonumber  \\
&+&(3\bar{\alpha}-2\sqrt{\bar{\alpha}}+2\,\bar{\alpha}\Lambda_\alpha)
\,\frac{\bar{k}^2}{4\bar{p}^{\frac{3}{2}}}
\, (\delta E^a_i \delta^i_a)^2\nn\\
&+&\bar{\alpha}^\frac{3}{2}(1+\Lambda_\alpha-2\Omega_\alpha)
\frac{1}{\bar{p}^{\frac{3}{2}}}W^{cdij}_{ab}\,( \partial_c \delta E^a_i) (\partial_d \delta E^b_j) ,\nn\\
\ea
\ba
\mathcal{H}^{(1)}_m&=&\bar{\beta}(1+\Lambda_\beta)\frac{\bar{\pi} }{\bar{p}^{\frac{3}{2}}}\delta \pi-\bar{\beta}(1+2\Lambda_\beta)\frac{\bar{\pi}^2}{4\bar{p}^{\frac{5}{2}}}\,\delta^i_a \delta E^a_i \nn\\
&+&\frac{1}{\beta(1+\Lambda_\beta)}\,\bar{p}^{\frac{3}{2}} V_{,\varphi}(\bar{\varphi}) \delta \varphi+\frac{\sqrt{\bar{p}}}{2}V(\bar{\varphi})
\delta^i_a\delta E^a_i ,\label{H1m}\nn\\~\\
\mathcal{H}^{(2)}_m&=& \bar{\beta}^2(1+\Lambda_\beta)^2\frac{1}{2\bar{p}^{\frac{3}{2}}}(\delta \pi)^2+\frac{V(\bar{\varphi})}{8\sqrt{\bar{p}}} (\delta^i_a \delta E^a_i )^2\nn\\
&+&\frac{1}{\bar{\beta}(1+\Lambda_\beta)^2}\frac{1}{2}\bar{p}^{\frac{3}{2}} V_{,\varphi\varphi}(\bar{\varphi}) (\delta \varphi)^2
\nonumber \\
&+& \bar{\beta}\left(2\bar{\beta}(1+\Lambda_\beta)^2-2\Lambda_\beta-1\right)
\frac{\bar{\pi}^2}{16\bar{p}^{\frac{7}{2}}}(\delta^i_a \delta E^a_i )^2\nn\\
&+&\bar{\beta}(1+2\Lambda_\beta)\frac{\bar{\pi}^2}{8\bar{p}^{\frac{7}{2}}}\delta^i_a \delta^j_b \delta E^a_j \delta E^b_i\nn\\
&-&\frac{V(\bar{\varphi})}{4\sqrt{\bar{p}}}\delta^i_a \delta^j_b \delta E^a_j \delta E^b_i+\frac{\sqrt{\bar{p}}}{2} V_{,\varphi}(\bar{\varphi})\delta \varphi(\delta^i_a \delta E^a_i) \nonumber\\
&-&\bar{\beta}^2(1+\Lambda_\beta)^2\frac{\bar{\pi} }{2\bar{p}^{\frac{5}{2}}}\delta \pi(\delta^i_a \delta E^a_i)\nn\\
&+&\frac{\bar{\alpha}^\frac{3}{2}}{\bar{\beta}}
\frac{(1+\Lambda_\alpha)}{(1+\Lambda_\beta)}\frac{1}{2} \sqrt{\bar{p}}\,(\partial^a \delta \varphi) \partial_a \delta \varphi .
\ea

 The total effective Hamiltonian on the spatially flat FRW background is expressed by
\ba
\textbf{H}^Q_{total}=\textbf{H}_{k=0}^Q[N]+\textbf{D}_{k=0}[N^a]
+\textbf{G}_{k=0}[\Lambda^i],\label{Htotal}
\ea
where both the diffeomorphism and Gauss constraint keep their classical forms.

\section{Cosmological perturbations on spatially flat FRW background}
In order to derive the possible observational effects of inverse-volume corrections, it is necessary to study the cosmological perturbations on the spatially flat FRW background. In this section, the main effort is dedicated to derive the equations of motion of gauge invariant perturbed variables.
\subsection{Canonical equations of motion}
In this article, we mainly focus on the effects of inverse-volume corrections during the slow-roll inflation. In the slow-roll period, it is reasonable to believe the backreaction of the perturbations is very weak such that the evolution of the homogeneous parts are not affected by the inhomogeneous parts; then, using the total Hamiltonian (\ref{Htotal}), the equations of motion for background variables are given by
\ba
\dot{\bar{k}} &=&-\,\bar{\alpha}(1-2\Lambda_\alpha)\frac{\bar{N}}{2\sqrt{\bar{p}}}\k^2
 -\bar{\beta}(1+2\Lambda_\beta)\frac{\kappa\bar{N}}{4\bar{p}^\frac{3}{2}}\bar{\pi}^2\nn\\
 &&+\frac{\kappa}{2} \sqrt{\bar{p}}\bar{N}V(\bar{\varphi}) , \label{dotk}  \\
\dot{\bar{p}} &=&  2\,\bar{\alpha}\bar{N} \sqrt{\bar{p}}\,\k , \label{dotp} \\
\dot{\bar{\varphi}} &=&  \bar{\beta} \frac{\bar{N}}{\bar{p}^{\frac{3}{2}}}\bar{\pi},  \label{dotvarphi} \\
\dot{\bar{\pi}} &=&  -\bar{N}  \bar{p}^{\frac{3}{2}} V_{, \varphi}(\bar{\varphi}) .
\ea

 By setting $\bar{N}=\sqrt{\p}$, we can get the quantum-corrected Friedmann, Raychaudhuri and Klein-Gorden equations in conformal time,
\ba
&&\mathcal{H}^2=\frac{\kappa}{3}\bar{\alpha}
\left(\frac{\bar{\varphi}'^{\,2}}{2\bar{\beta}}+\p V(\bar{\varphi})\right),\label{Friedmann}\\
&&\mathcal{H}'=(1-\Lambda_\alpha)\,\mathcal{H}^2
-\frac{\kappa}{2}\frac{\bar{\alpha}}{\bar{\beta}}(1+\Lambda_\beta)
\,\bar{\varphi}'^{\,2},\label{Raychaudhuri}\\
&&\bar{\varphi}''+2(1+3\Lambda_\beta)\,\mathcal{H}+\bar{\beta}\p V_{,\varphi}(\bar{\varphi})=0,\label{Klein-Gorden}
\ea
where the Hubble rate $\mathcal{H}\equiv\frac{p'}{2p}$ and the prime denotes a derivative with respect to the conformal time $\eta$.

The quantum-corrected Hamilton' equation for the perturbed variables $\delta K_a^i$, $\delta E^a_i$, $\delta \varphi$ and $\delta \pi$ are separately written as
\ba
\dot{\delta K^i_a}&=&\bar{k}\,(\partial_a \delta N^c)\delta_c^i+\frac{2}{\sqrt{\bar{p}}} \bar{\alpha}\,(1+\Lambda_\alpha)(\partial_a \partial^i \delta N)\nn\\
&&+\frac{\delta^i_a\delta N}{2\sqrt{\bar{p}}}\left( -\bar{\beta}(1+2\Lambda_\beta)\frac{\kappa\bar{\pi}^2}{2\bar{p}^2}
+\kappa \bar{p} V(\bar{\varphi})\right)\nonumber \\
&&-\frac{\delta^i_a\delta N}{2\sqrt{\bar{p}}}(3\bar{\alpha}-2\sqrt{\bar{\alpha}})
\k^2-\frac{\bar{N}}{\sqrt{p}}(2\bar{\alpha}-\sqrt{\bar{\alpha}}) \k\,\delta K_a^i\nn \\
&&-\frac{\bar{N}}{2\bar{p}^{\frac{3}{2}}}
(3\bar{\alpha}-2\sqrt{\bar{\alpha}}+2\,\bar{\alpha}\Lambda_\alpha)
\k^2(\delta_a^j\delta_c^i\delta E^c_j)\nn \\
&&+\frac{\bar{N}}{4\bar{p}^{\frac{3}{2}}}
(3\bar{\alpha}-2\sqrt{\bar{\alpha}}+2\,\bar{\alpha}\Lambda_\alpha)
\k^2(\delta_c^j\delta E^c_j)\delta^i_a\nn \\
&&-\frac{\bar{N}}{2 \bar{p}^\frac{3}{2}}\bar{\alpha}^\frac{3}{2}(1+\Lambda_\alpha-2\Omega_\alpha)\, \delta^{ij} W^{cdij}_{ab}\partial_c \partial_d \delta E^b_j\nn \\
&&-\frac{\bar{N}}{2 \bar{p}^\frac{3}{2}}\bar{\alpha}^\frac{3}{2}(1+\Lambda_\alpha-2\Omega_\alpha)\, \delta^{ij} W^{cdji}_{ba}\partial_d \partial_c \delta E^b_j \nn \\
&&+\frac{\bar{N}}{2\bar{p}^{\frac{3}{2}}}\left(\bar{\beta}(1+2\Lambda_\beta)
\frac{\kappa\bar{\pi}^2}{2\bar{p}^2}-\kappa \bar{p}V(\bar{\varphi})\right)(\delta_a^j\delta_c^i\delta E^c_j)\nn\\
&&+\frac{\bar{N}\kappa\pi^2}{8\bar{p}^{\frac{7}{2}}}
\bar{\beta}\left(2\bar{\beta}(1+\Lambda_\beta)^2-2\Lambda_\beta-1\right)
(\delta_c^j\delta E^c_j)\delta_a^i\nn\\
&&+\frac{\bar{N}\kappa V(\bar{\varphi})}{4\sqrt{\bar{p}}}(\delta_c^j\delta E^c_j)\delta_a^i-\frac{\kappa \bar{N} }{2\bar{p}^{\frac{5}{2}}}\bar{\beta}^2(1+\Lambda_\beta)^2\pi \delta\pi\delta_a^i\nn\\
&&+\frac{\sqrt{\bar{p}}\kappa \bar{N} \delta_a^i}{2}V_{,\varphi}(\bar{\varphi})\delta \varphi+\k\,\epsilon_{kj}{}^i\delta_a^j\delta\Lambda^k , \label{dotdeltak}
\ea
\ba
\dot{\delta E^a_i}&=&-\bar{p} \left(\partial_i \delta N^a -\delta^a_i\partial_c \delta N^c\right)+2\sqrt{\alpha}\sqrt{\bar{p}}\,\k\,\delta^a_i\delta N\nn\\
&&+\sqrt{\alpha}\bar{N} \sqrt{\bar{p}}\,(\delta^c_j\delta K^j_c )\delta^a_i
-\sqrt{\alpha}\bar{N}\sqrt{\bar{p}}(\delta^c_i \delta^a_j\delta K^j_c) \nn\\ &&+\frac{N}{\sqrt{\bar{p}}}\left(2\alpha-\sqrt{\alpha}\right)\k\delta E^a_i-\p\,\epsilon_{ki}{}^j\delta^a_j\delta\Lambda^k,\label{dotdeltaE}\\
\dot{\delta\varphi}&=&\bar{\beta}(1+\Lambda_\beta)\frac{\pi\delta N}{\bar{p}^{\frac{3}{2}}}+\bar{\beta}^2(1+\Lambda_\beta)^2\frac{\bar{N}\delta \pi }{\bar{p}^{\frac{3}{2}}}\nn\\
&&-\bar{\beta}^2(1+\Lambda_\beta)^2
\frac{\bar{N}\bar{\pi}}{2\bar{p}^{\frac{5}{2}}}(\delta_a^i\delta E^a_i) ,\label{dotdeltaphi}\\
\dot{\delta \pi}&=&\bar{\pi}\partial_a\delta N^a-\frac{1}{\bar{\beta}(1+\Lambda_\beta)}\bar{p}^{3/2} V_{,\varphi}(\bar{\varphi})  \delta N\nn\\
&&-\frac{1}{\bar{\beta}(1+\Lambda_\beta)^2}\bar{N}\bar{p}^{\frac{3}{2}}
V_{,\varphi\varphi}(\bar{\varphi})\delta\varphi
-\frac{\bar{N}\bar{p}^{\frac{1}{2}}}{2}V_{,\varphi}(\bar{\varphi})\delta_a^i\delta E^a_i\nonumber\\
&&+\frac{\bar{\alpha}^\frac{3}{2}}{\bar{\beta}}
\frac{(1+\Lambda_\alpha)}{(1+\Lambda_\beta)}\bar{N}\sqrt{\bar{p}}
\,\delta^{ab}(\partial_b\partial_a\delta \varphi).\label{dotdeltapi}
\ea

\subsection{Gauge invariant variables}
Due to the quantum corrections, the gauge transformations of perturbations behave differently from the classical theory, thus, it is necessary to reconstruct the gauge invariant variables first.

 To begin with, we decompose the configuration variable into the scalar, vector and tensor modes, i.e.
\ba
\delta E^a_i&=&\bar{p}\Big[-2\psi\delta^a_i+(\delta^a_i\partial^d\partial_d-\partial^a \partial_i)E-\frac{1}{2}\partial^aF_i-\frac{1}{2}\partial_iF^a\nonumber\\
&&\quad-\frac{1}{2}h^a_{~i}\Big] ,
\end{eqnarray}
where to guarantee that the physical variables are invariant under Gauss transformation we set $h^a_{~i}=h_{~a}^{i}$ and put the same prefactor $\frac{1}{2}$ before the vector modes \cite{Han:2017}. In addition, the perturbed lapse function and shift vector are decomposed as
\ba
\delta N=\bar{N}\phi ,\qquad\delta N^a=\partial^a B+S^a ,
\ea
then, from  Eq. (\ref{dotdeltaE}), we find the internally gauge invariant momentum $\delta K_{a}^{i}$ can be decomposed as
\begin{eqnarray}
&&\sqrt{\bar{\alpha}}\,\delta K_{a}^i\nn\\
&&=-\delta^a_i\left[\psi'+\frac{\mathcal{H}}{\sqrt{\bar{\alpha}}}(\psi
+\phi)\right]+\partial_a\partial^i\left[\frac{\mathcal{H}}{\sqrt{\bar{\alpha}}}E
-(B-E')\right]\nonumber\\
&&\quad+\frac{1}{2}\bigg[\frac{\mathcal{H}}{\sqrt{\bar{\alpha}}}
(\partial_aF^i+\partial^iF_a)
+(\partial_aF^i+\partial^iF_a)'\nn\\
&&\qquad\quad-(\partial_aS^i+\partial^iS_a)\bigg]+\frac{1}{2}\bigg[{h^{i}_{~a}}'
+\frac{\mathcal{H}}{\sqrt{\bar{\alpha}}}h^{i}_{~a}\bigg],\label{decomdeltak}
\end{eqnarray}
where $h^{i}_{~a}\equiv\delta^i_b\delta^j_ah^{b}_{~j}$. Moreover, from (\ref{dotdeltaphi}) the decomposition of $\delta\pi$ reads
\ba
\delta\pi=\frac{\p\,\delta\varphi'}{\bar{\beta}^2(1+\Lambda_\beta)^2}
-\frac{\p\,\bar{\varphi}'}{\bar{\beta}}\left(3\psi-\Delta E+\frac{\phi}{\bar{\beta}(1+\Lambda_\beta)}\right),\label{deltapi}\nn\\
\ea
where $\Delta\equiv\delta^{ab}\partial_a\partial_b$.

According to Ref. \cite{Bojowald:2008b}, under the small coordinate transformation parametrized by
\ba
&&x^{\mu}\rightarrow x^{\mu}+\xi^{\mu},\nn\\
&&\xi^{\mu}\equiv(\xi_0,\xi^{a}),\quad\xi^a\equiv\partial^a\xi+\tilde{\xi}^a,
\ea
where $\partial_a\tilde{\xi}^a=0$, the infinitesimal change of a perturbed phase- space variable $X$  is given by
\ba
\delta_{[\xi_0, \xi^a]}X\equiv\{X, \textbf{H}^{(2)}[\bar{N}\xi_0]+\textbf{D}[\xi^a]\} ,\label{dtransX}
\ea
where
\ba
\textbf{H}^{(2)}[\bar{N}\xi_0]\equiv \int_\mathcal{C} d^3x \bar{N}\xi_0 \left[\frac{1}{2 \kappa}\mathcal{H}^{(1)}_g+\mathcal{H}^{(1)}_m \right],\label{H(2)}
\ea
 in which the expressions of $\mathcal{H}^{(1)}_g$ and $\mathcal{H}^{(1)}_m$ are given in (\ref{H1g}) and (\ref{H1m}). With the help of (\ref{H(2)}), it is easy to find the gauge transformation of $X$'s time derivative,
\ba
\delta_{[\xi_0, \xi^a]}\dot{X}=\dot{(\delta_{[\xi_0, \xi^a]} X)}-\bar{\alpha}^\frac{3}{2}(1+\Lambda_\alpha)\,\delta_{[0, \partial^a\xi_0]}X .\nn\\\label{dtransdotX}
\ea

In what follows we separately consider the different modes. Let us start with the scalar modes of perturbation. Using the auxiliary expression (\ref{dtransX}) we find
\ba
&&\delta_{[\xi_0, \xi^a]}(\sqrt{\bar{\alpha}}\delta K_a^i)\nn\\
&&=-\frac{1}{2}\frac{\mathcal{H}^2}{\sqrt{\bar{\alpha}}}
(3-\frac{2}{\sqrt{\bar{\alpha}}})\,\xi_0\delta^i_a\nn\\
&&~~~+\frac{\kappa}{2}\sqrt{\bar{\alpha}}
\left[-\frac{\bar{\varphi}'^2}{2\bar{\beta}}(1+2\Lambda_\beta)+\p V(\bar{\varphi})\right]\,\xi_0\delta^i_a\nn\\
&&~~~+\partial_a\partial^i
\left[\frac{\mathcal{H}}{\sqrt{\bar{\alpha}}}\xi
+\bar{\alpha}^\frac{3}{2}(1+\Lambda_\alpha)\xi_0\right],\\
&&\delta_{[\xi_0, \xi^a]}\delta E^a_i=2\frac{\mathcal{H}}{\sqrt{\bar{\alpha}}}\p\,\xi_0\delta^i_a
+\p(\delta^a_i\Delta\xi-\partial_a\partial^i\xi),\\
&&\delta_{[\xi_0, \xi^a]}\delta\varphi=\bar{\varphi}'(1+\Lambda_\beta)\xi_0,\\
&&\delta_{[\xi_0, \xi^a]}\delta\pi=\frac{\p\,\bar{\varphi}'}{\bar{\beta}}\Delta\xi
-\frac{\p^2V_{,\varphi}(\varphi)}{\bar{\beta}(1+\Lambda_\beta)}\xi_0.
\ea

 From the above equations along with (\ref{dtransdotX}), we obtain
 \ba
 &&\delta_{[\xi_0, \xi^a]}\psi=-\frac{\mathcal{H}}{\sqrt{\alpha}}\xi_0,\quad
 \delta_{[\xi_0, \xi^a]}\phi=\mathcal{H}\xi_0+{\xi_0}',\nn\\
 &&\delta_{[\xi_0, \xi^a]}(B-E')=-\bar{\alpha}^\frac{3}{2}(1+\Lambda_\alpha)\xi_0.
 \ea
 It is then not difficult to show that the following combinations of perturbations are gauge invariant:
 \ba
\Psi&\equiv&\psi-\frac{\mathcal{H}(B-E')}{\bar{\alpha}^2(1+\Lambda_\alpha)},\label{Psi}\\
\Phi&\equiv&\phi+\left(\frac{B-E'}{\bar{\alpha}^\frac{3}{2}(1+\Lambda_\alpha)}\right)'
+\frac{\mathcal{H}(B-E')}{\bar{\alpha}^\frac{3}{2}(1+\Lambda_\alpha)},\label{Phi}\\
\delta\tilde{\varphi}&\equiv&\delta\varphi+
\frac{1+\Lambda_\beta}{\bar{\alpha}^\frac{3}{2}(1+\Lambda_\alpha)}\bar{\varphi}'(B-E').
\label{titphi}
 \ea
 Obviously,  $\Phi$ and $\Psi$ reproduce the Bardeen potential in the classical limit.

  For vector perturbation, by simply repeating the procedure above, we find the quantity
 \be
 V^a\equiv S^a-{F^a}'\label{givector}
 \ee
 is invariant under coordinate transformations.

\subsection{Gauge invariant linear perturbations}
Using the gauge invariant variables defined above, we proceed to derive the corresponding equations of motion of these variables. The derivations in the following are straightforward but the calculations are quite tedious. For brevity, at some steps we will skip the details and directly give the results.
\subsubsection{Gauge invariant scalar perturbations}
First, by varying the diffeomorphism constraint with respect to the shift vector, we find the perturbed diffeomorphism constraint equation:
\ba
\kappa\frac{\delta \textbf{D}_{k=0}[N^a]}{\delta(\delta N^a)}&=&2\bar{p}\,\delta^b_i \partial_{[a}\delta K^i_{b]}
-\bar{k}\,\,\delta^i_a \partial_b \delta E^b_i+\kappa \bar{\pi}\partial_a\delta\varphi\nn\\
&=&0.
\ea
 Using the definition in (\ref{Psi}), (\ref{Phi}) and (\ref{titphi}) and the background evolution equations (\ref{Friedmann})~(\ref{Klein-Gorden}), it translates into
\ba
\partial_a\left(\Psi'+\frac{\mathcal{H}}{\sqrt{\bar{\alpha}}}\Phi\right)=
\frac{\kappa}{2}\frac{\sqrt{\bar{\alpha}}}{\bar{\beta}}\bar{\varphi}'
\partial_a\delta\tilde{\varphi}.\label{Dconspsi}
\ea
Similarly, variation of the Hamiltonian constraint with respect to the perturbed lapse function gives
\ba
&&\frac{1}{2\kappa}\Big[-4\,\sqrt{\bar{\alpha}}\sqrt{\bar{p}}\,\bar{k}\,\delta^a_i\delta K^i_a
-(3\bar{\alpha}-2\sqrt{\bar{\alpha}})\frac{\bar{k}^2}{\sqrt{\bar{p}}}
\,\delta^i_a\delta E^a_i\nn\\
&&~~~+\bar{\alpha}\,(1+\Lambda_\alpha)\frac{2}{\sqrt{\bar{p}}}\,\partial_a \partial^i \delta E^a_i\Big]+\bar{\beta}(1+\Lambda_\beta)\frac{\bar{\pi} }{\bar{p}^{\frac{3}{2}}}\delta \pi\nn\\
&&~~~-\bar{\beta}(1+2\Lambda_\beta)\frac{\bar{\pi}^2}{4\bar{p}^{\frac{5}{2}}}\,\delta^i_a \delta E^a_i
+\frac{1}{\beta(1+\Lambda_\beta)}\,\bar{p}^{\frac{3}{2}} V_{,\varphi}(\bar{\varphi}) \delta \varphi\nn\\
&&~~~+\frac{\sqrt{\bar{p}}}{2}V(\bar{\varphi})
\delta^i_a\delta E^a_i =0,
\ea
of which the scalar part can reduce to the gauge invariant version:
\ba
&&\bar{\alpha}(1+\Lambda_\alpha)\Delta\Psi-3\frac{\mathcal{H}}{\bar{\alpha}}
(\Psi'+\frac{\mathcal{H}}{\sqrt{\bar{\alpha}}}\Phi)\nn\\
&=&\frac{\kappa}{2}\frac{1}{\bar{\beta}^2(1+\Lambda_\beta)}
\left[\bar{\varphi}'\delta\tilde{\varphi}'
-(1+\Lambda_\beta)\bar{\varphi}'^2\Phi
+\bar{\beta}\p V_{,\varphi}(\bar{\varphi})\delta \tilde{\varphi}\right].\label{Hconspsi}\nn\\
\ea

Next, using (\ref{decomdeltak}), it is direct to show that the diagonal parts of (\ref{dotdeltak}) yield the following equation,
\ba
&&\Psi''+\mathcal{H}\Big[(2+\Lambda_\alpha)\Psi'+\frac{\Phi'}{\sqrt{\bar{\alpha}}}\Big]
+\frac{1}{\sqrt{\bar{\alpha}}}\Big[\mathcal{H}'+2\mathcal{H}^2(1+\Lambda_\alpha)\Big]\Phi
\nn\\
&&=\frac{\kappa}{2}\sqrt{\bar{\alpha}}\Big[\frac{\bar{\varphi}'}{\bar{\beta}}
\delta\tilde{\varphi}'-\p V_{,\varphi}(\bar{\varphi})\delta\tilde{\varphi}\Big],\label{eqpsi}
\ea
and the off-diagonal part gives
\ba
(1+\Lambda_\alpha)\Phi=\sqrt{\bar{\alpha}}(1+\Lambda_\alpha-2\Omega_\alpha)\Psi.
\label{conspsiphi}
\ea
Moreover, with the help of (\ref{dotdeltapi}) and (\ref{deltapi}), the Klein-Gordon equation can also be reformulated in the gauge invariant form,
\ba
&&\delta\tilde{\varphi}''+\bigg[2\mathcal{H}
-2\frac{\big(\bar{\beta}(1+\Lambda_\beta)\big)'}{\bar{\beta}(1+\Lambda_\beta)}\bigg]
\delta\tilde{\varphi}'\nn\\
&&\quad-\bar{\alpha}^\frac{3}{2}\bar{\beta}(1+\Lambda_\beta)
(1+\Lambda_\alpha)\Delta\delta\varphi+\bar{\beta}\p V_{,\varphi\varphi}(\bar{\varphi})
\delta\tilde{\varphi}\nn\\
&&\quad+\bigg[2\bar{\beta}(1+\Lambda_\beta)\p V_{,\varphi}(\bar{\varphi})
+\frac{\bar{\varphi}'}{\bar{\beta}}\big(\bar{\beta}(1+\Lambda_\beta)\big)'\bigg]\Phi
\nn\\
&&\quad-\left[(1+\Lambda_\beta)\bar{\varphi}'+3\bar{\beta}(1+\Lambda_\beta)^2\bar{\varphi}'\right]\Psi'=0
.\label{eqdelvarphi}\ea

Finally, from the Eqs.  (\ref{Dconspsi})~(\ref{eqdelvarphi}) along with background equations of motion, after a long calculation, we obtain the quantum-corrected Mukhanov equation of density perturbation,
\ba
v''-c_s^2\Delta v-\frac{z_s''}{z_s}v= 0, \label{SMukheqn}
\ea
where
\ba
v&\equiv&\sqrt{\p}\left(\frac{\delta\tilde{\varphi}}{\bar{\beta}(1+\Lambda_\beta)}
+\frac{\sqrt{\bar{\alpha}}}{\bar{\beta}}\frac{\bar{\varphi}'}{\mathcal{H}}\Psi\right),\\
c_s^2&\equiv&\bar{\alpha}^\frac{3}{2}\bar{\beta}(1+\Lambda_\beta)
(1+\Lambda_\alpha),\label{cs}\\
z_s&\equiv&\sqrt{\p}\frac{\sqrt{\bar{\alpha}}}{\bar{\beta}}\frac{\bar{\varphi}'}{\mathcal{H}}.
\ea
 Notice that not only the definition of Mukhanov variable but also the propagation speed of density perturbation is modified by the inverse-volume corrections.
\subsubsection{Gauge invariant tensor perturbations}
Since the tensor mode does not change under coordinate transformations, we can directly obtain its gauge invariant equations of motion from (\ref{dotdeltak}):
\ba
{h^{i}_{~a}}''+(2+\Lambda_\alpha)\mathcal{H}{h^{i}_{~a}}'
-\bar{\alpha}^2(1+\Lambda_\alpha-2\Omega_\alpha)
\Delta h^{i}_{~a}=0.\nn\\
\ea
Equivalently, it can be written in the form of the tensor Mukhanov equation,
\ba
h''-c_t^2\Delta h-\frac{z_t''}{z_t}h= 0, \label{TMukheqn}
\ea
where
\ba
h&\equiv&\frac{\sqrt{\p}}{\sqrt{2\kappa}\bar{\alpha}^\frac{1}{4}}h^{i}_{~a},\\
c_t^2&\equiv&\bar{\alpha}^2(1+\Lambda_\alpha-2\Omega_\alpha),\label{ct}\\
z_t&\equiv&\frac{\sqrt{\p}}{\bar{\alpha}^\frac{1}{4}}.
\ea
Comparing it with (\ref{cs}), we observe that the propagation speed of tensor perturbations is different from that of density perturbations.
\subsubsection{Gauge invariant vector perturbations}
For gauge invariant vector perturbations, we define
\be
\Sigma^a_i\equiv\frac{1}{2}(\partial_iV^a+\partial^aV_i) ,
\ee
where the expression of $V^a$ is given in (\ref{givector}). Since the perturbation of scalar field does not generate the vector modes, the gauge invariant equation can be derived directly from (\ref{dotdeltak}),
\be
{\Sigma^a_i}'+(2+\Lambda_\alpha)\mathcal{H}\Sigma^a_i=0 ,\label{eomSigma}
\ee
which shows that the cosmological vector mode decays more quickly than the classical scenario provided $\Lambda_\alpha>0$.  Interestingly, for scalar matter, Eq.  (\ref{eomSigma}) coincides with the Eq. (39) in Ref. \cite{Bojowald:2007a}, although the latter was obtained following a somewhat different consideration.
\section{Spectral indices with quantum corrections}
In this section, we aim to solve the Mukhanov equation associated with the scalar and tensor modes; then, we derive the inverse-volume- corrected spectral indices and the tensor-to-scalar ratio. For simplicity, we only keep the leading order quantum corrections in the final results.

In the weak quantum gravity regime such as the slow-roll inflation, the inverse-volume correction functions are often expanded as
\ba
\bar{\alpha}(\p)&=&1+\alpha_0\zeta(\p),\nn\\
\bar{\beta}(\p)&=&1+\beta_0\zeta(\p),
\ea
where
\ba
\zeta(\p)\equiv\left(\frac{l^2_{Pl}}{\bar{p}\mathcal{V}^{\frac{2}{3}}_0}\right)^\sigma \label{qgcor}
\ea
represents the effective correction which varies with respect to the scale factor.  The constants $\alpha_0$, $\beta_0$ and $\sigma$ mainly depend on specific choice of parametrization in LQC. In some viable models, the coefficients $\alpha_0$ and $\beta_0$ are estimated to be the order of $O(10^{-1})$, and $\sigma$ is a positive number lying in the range $0<\sigma\leq3$; the fraction in (\ref{qgcor}) stands for the ratio of Planck length compared to the quantum gravity length scale \cite{Bojowald:2010}. Note that in the anomaly-free algebra these constants are subjected to another restriction, i.e. the Eq.(\ref{consiscon}), from which we infer
 \ba
 \beta_0(2\sigma-3)(\sigma+3)=-\frac{9}{2}\alpha_0,\label{alphabeta}
 \ea
 where in the derivation we ignore the terms proportional to $\zeta^2$, and the same will be done in the following.

\subsection{Scalar spectrum and spectral index }
In momentum space, the Mukhanov equation (\ref{SMukheqn}) becomes
\ba
v_k''+\left(c_s^2 k^2 -\frac{z_s''}{z_s}\right)v_k= 0, \label{SMukheqn}
\ea
with
\ba
c_s^2&=&1+\left[(\frac{3}{2}+\sigma)\alpha_0+(1+\frac{\sigma}{3})\beta_0\right]\zeta\nn\\
&=&1+\frac{6-2\sigma^2}{3-2\sigma}\,\alpha_0\zeta,\label{cs2}\\
z_s&=&\sqrt{\p}\frac{\bar{\varphi}'}{\mathcal{H}}
\left[1+(\frac{\alpha_0}{2}-\beta_0)\zeta\right]\nn\\
&=&\sqrt{\p}\frac{\bar{\varphi}'}{\mathcal{H}}
\left[1+\frac{2\sigma^2+3\sigma}{2(2\sigma-3)(\sigma+3)}\,\alpha_0\zeta\right],\label{zs2}
\ea
where in the last step of (\ref{cs2}) and (\ref{zs2}) we have used Eq. (\ref{alphabeta}).

 From (\ref{cs2}) we see that the propagation speed $c_s$ will be larger than 1 when $\sigma$ takes values in the range $0<\sigma<\frac{3}{2}$ and $\sqrt{3}<\sigma\leq3$ for $\alpha_0>0$; thus, for most permissible values of $\sigma$, the propagation speed
 will become superluminal, violating the law of causality. Nevertheless, one should note that the physical speed of light is also affected by the quantum gravity corrections.
As discussed in Ref. \cite{Bojowald:2007b}, the group velocity of electromagnetic wave propagation is
\ba
v_{EM}=\sqrt{\alpha_{EM}\beta_{EM}},
\ea
where $\alpha_{EM}(E^a_i)$ and $\beta_{EM}(E^a_i)$ separately represent the inverse-volume corrections before the electric field part and magnetic field part in the Hamiltonian constraint and satisfy $\alpha_{EM}\rightarrow1$, $\beta_{EM}\rightarrow1$ in the classical limit. In the anomaly-free algebra, both $\alpha_{EM}$ and $\beta_{EM}$ are related to the correction function $\alpha(E^a_i)$ and some counterterms in the gravitational Hamiltonian constraint. By solving the complex anomaly equations we can express the counterterms as functions of $\alpha$, $\alpha_{EM}$ and $\beta_{EM}$, which is still a open problem yet. It is only after  $\alpha_{EM}$ and $\beta_{EM}$ are expressed as unambiguous functions of $\alpha$ that we can compare the physical speed of light with that of density perturbations or gravitational waves.

The scalar power spectrum at the horizon crossing is given by
\ba
P_s\equiv\frac{k^3}{2\pi^2}\Big|\frac{v_k}{z_s}\Big|^2\bigg|_{k|\eta|=1}.\label{sPs}
\ea

Following the approach developed in Ref. \cite{Bojowald:2010} and using (\ref{cs2}) and (\ref{zs2}), the mode function at the horizon crossing can be expressed as
\ba
|v_k|^2=\frac{1}{2k}\left[1-\frac{(6-2\sigma^2)\,\alpha_0\zeta}{(2\sigma+1)(3-2\sigma)}\right]
\bigg|_{k|\eta|=1}.\label{v2}
\ea
Defining the slow roll parameters
\ba
\epsilon&\equiv&1-\frac{\mathcal{H}'}{\mathcal{H}^2},\nn\\
\delta&\equiv&1-\frac{\bar{\varphi}''}{\mathcal{H}\bar{\varphi}'},
\ea
we have
\ba
z^2_s=\frac{2\p}{\kappa}
\left[\epsilon-\left(\sigma+\frac{3\epsilon}{6-4\sigma}\right)\,\alpha_0\zeta\right].
\label{zs3}
\ea
Plugging the identity (\ref{v2}) and (\ref{zs3}) into (\ref{sPs}) and then using
$k=\mathcal{H}$ at the horizon crossing, we get
\ba
P_s=\frac{\kappa}{8\pi^2}\frac{\mathcal{H}^2}{\p\,\epsilon}
\left[1+\left(\frac{4\sigma^2+6\sigma-9}{(2\sigma+1)(6-4\sigma)}
+\frac{\sigma}{\epsilon}\right)\alpha_0\zeta\right],\label{sPs2}\nn\\
\ea
which yields the scalar spectral index following the standard procedure,
\ba
n_s-1=-4\epsilon+2\delta+\lambda_s\alpha_0\zeta,
\ea
where
\ba
\lambda_s\equiv\frac{-12\sigma^4-28\sigma^3+6\sigma^2+36\sigma}{(\sigma+3)(2\sigma+1)(3-2\sigma)},
\ea
such that for the range $0<\sigma\leq1.02$, the quantum corrections will increase the scalar spectral index.
\subsection{Tensor spectral index and tensor-to-scalar ratio }
The square of speed for gravitational waves follows from (\ref{ct}),
\ba
c^2_t=1+(2-2\sigma^2-\sigma)\,\alpha_0\zeta.
\ea
The tensor spectrum turns out
\ba
P_t&\equiv&\frac{4\kappa k^3}{\pi^2}\Big|\frac{h_k}{z_t}\Big|^2\bigg|_{k|\eta|=1}\nn\\
&=&\frac{2\kappa}{\pi^2}\frac{\mathcal{H}^2}{\p}
\left(1+\frac{(2\sigma-1)(2\sigma+3)}{2(2\sigma+1)}\alpha_0\zeta\right),\label{tPs}
\ea
which leads to
\ba
n_t=-2\epsilon+\lambda_t\alpha_0\zeta,
\ea
where
\ba
\lambda_t\equiv\frac{\sigma(1-2\sigma)(2\sigma+3)}{(2\sigma+1)}.
\ea
From (\ref{sPs2}) and (\ref{tPs}) one can find the tensor-to-scalar ratio,
\ba
r&\equiv&\frac{P_t}{P_s}=16\epsilon
\left[1+\left(\frac{6\sigma-4\sigma^3}{(2\sigma+1)(3-2\sigma)}
-\frac{\sigma}{\epsilon}\right)\alpha_0\zeta\right],\nn\\
\ea
from which we see that, due to the small value of $\epsilon$, the inverse-volume corrections tend to lower the tensor-to-scalar ratio, making it more difficult to detect.
\section{Conclusion and Remarks}
  In this article, the issue of anomaly freedom of perturbative LQC with inverse-volume corrections is revisited. Now we summarize the basic idea and what has been achieved in this paper.

Despite the considerable progress made in the anomaly-free approach so far, there remains some problems which bother the effective perturbation theory. One of the key problems is that the counterterms in the effective perturbed constraints cannot be uniquely fixed for inverse-volume corrections on spatially flat FRW background, which causes ambiguities in the subsequent cosmological perturbations and weakens the theory's predictive power. In this article, without any additional input in the anomaly-free approach, we find that after including the positive spatial curvature each counter term can be uniquely expressed in terms of the inverse-volume correction functions and their expressions agree with the results obtained in spatially flat case \cite{Cailleteau:2013}, thus, in the large $r_o$ limit, the Hamiltonian density on the spatially closed background successfully reduces to the Hamiltonian density on the spatially flat background with each component having fixed expressions. In the latter part of this paper, by using the anomaly-free Hamiltonian, we derived the gauge invariant cosmological perturbations and carried out a preliminary study of the observational quantum effects on the spatially flat background.  Before we end this paper, several remarks are in order.

 First, despite the quantization ambiguities, in the anomaly-free algebra the study of inverse-volume corrections are perhaps more reliable compared with the study of holonomy corrections because the current study of the holonomy-corrected perturbed constraint does not include the contributions from the higher order spatial derivatives of the connection  which could be comparable to the contributions from higher powers of extrinsic curvature when the quantum gravity effect becomes strong, whereas this problem does not appear for inverse-volume corrections, thus, at present, the issue of anomaly freedom for inverse-volume corrections should be more seriously taken than the holonomy corrections.

 Second, from (\ref{cs2}), there seems a danger caused by the superluminal propagation speed of density perturbations for most admissible values of $\sigma$. However, one must compare it with the physical speed of light which can be fixed by anomaly-free algebra including both the electromagnetic fields and the scalar field; this task needs to be done in future research. Moreover, it is interesting to observe that the propagation speed for density and tensor perturbations are different for inverse-volume corrections, whereas in the case of holonomy corrections, the speeds are the same for both and always smaller than 1 with minimally coupled ordinary matter.

 Finally, it is worth mentioning the following. 1) Although we only focus on the  slow-roll inflation in which the quantum gravity effects are weak, the results obtained in Secs. III and IV also apply to the deep quantum gravity regime such as the preinflationary period  because the algebra is still closed in that regime; thus,  we can perform the research for the preinflationary quantum gravity effects by using  different initial conditions.
 2) For simplicity, the calculations in Sec. V follow the approach developed in Ref. \cite{Bojowald:2010}; however, when obtaining the mode functions at the horizon crossing, some approximations used there can lead to non-negligible errors. A corresponding more precise and much involved treatment is given in Ref. \cite{Zhu:2015}; using the methods there, combined with the results in this article, we can perform a more accurate analysis in the context of concrete models with various scalar potentials. The detailed work is left for future study.

\begin{acknowledgments}
The author thanks Dr. Long Chen for helpful discussions. This work is supported by NSFC under Grant No.11647068 and Nanhu Scholars Program for Young Scholars of Xinyang Normal University.
\end{acknowledgments}
\appendix
\section{Calculation of the bracket (\ref{formula1}) }
\ba
&&\bigg\{\int_\mathcal{C} d^3x\frac{\delta N}{\sqrt{\p}}\,\, \eb^a_j\bar{D}_a \bar{D}_b \delta E^b_j ,\,\int_\mathcal{C}  d^3x\delta N^a\bar{p}\sqrt{\q}\,\,\eb^b_i \bar{D}_{[a}\delta K^i_{b]}\bigg\}\nn\\
=&&\bigg\{\int_\mathcal{C}  d^3x\frac{\delta E^b_j}{\sqrt{\p}}\, \eb^a_j\bar{D}_a \bar{D}_b
\delta N,\,\int_\mathcal{C}  d^3x\sqrt{\q}\p\,\,\eb^b_i\delta K^i_{[a}\bar{D}_{b]}\delta N^a\bigg\}\nn\\
=&&\frac{\kappa}{2}\int_\mathcal{C} d^3x\sqrt{\q}\sqrt{\p}\,\,\eb^b_i\,\eb^c_i(\bar{D}_a\delta N^a)(\bar{D}_c \bar{D}_b \delta N)\nn\\
&&-\frac{\kappa}{2}\int_\mathcal{C} d^3x\sqrt{\q}\sqrt{\p}\,\,\eb^b_i\,\eb^c_i(\bar{D}_c\delta N^a)(\bar{D}_b \bar{D}_a \delta N)\nn
\ea
\ba
=&&\kappa\int_\mathcal{C}  d^3x\sqrt{\q}\sqrt{\p}\,\,\eb^b_i\,\eb^c_i
\left[(\bar{D}_{[a}\delta N^a)\bar{D}_{c]} \bar{D}_b \delta N\right]\nn\\
=&&\kappa\int_\mathcal{C}  d^3x\sqrt{\q}\sqrt{\p}\,\delta N\,\eb^b_i\,\eb^c_i
\left(\bar{D}_b\bar{D}_{[c} \bar{D}_{a]} \delta N^a\right)\nn\\
=&&-\frac{\kappa}{2}\int_\mathcal{C} d^3x\sqrt{\q}\sqrt{\p}\,\delta N\,\eb^b_i\,\eb^c_i\bar{D}_b
\left(\bar{D}_{a} \bar{D}_{c} \delta N^a-\bar{D}_{c} \bar{D}_{a}\delta N^a\right)\nn\\
=&&-\frac{\kappa}{2}\int_\mathcal{C}  d^3x\sqrt{\q}\sqrt{\p}\,\delta N\,\eb^b_i\,\eb^c_i\bar{D}_b (R^{(3)}_{ac}\delta N^a),\label{bracketapp}
\ea
where $R^{(3)}_{ac}$ denotes the three-dimensional Ricci tensor; on 3-sphere, it is proportional to the fiducial metric£º
\ba
R^{(3)}_{ac}=\frac{2}{r^2_o}\q_{ac}.\label{3Ricci}
\ea
where  $r_o$ denotes the radius of $\Sigma$ with respect to the fiducial metric. Plugging (\ref{3Ricci}) into (\ref{bracketapp}), we have
\ba
&&-\frac{\kappa}{2}\int_\mathcal{C}  d^3x\sqrt{\q}\sqrt{\p}\,\delta N\,\eb^b_i\,\eb^c_i\bar{D}_b (R^{(3)}_{ac}\delta N^a)\nn\\
&&=\kappa\int_\mathcal{C}d^3x \sqrt{\q}\sqrt{\bar{p}}\,\,\frac{1}{r_o^2}\,\delta N^a \bar{D}_a \delta N,
\ea
which is exactly the result on the right-hand side of (\ref{formula1}).

\end{document}